\documentclass[%
 reprint,
showpacs,preprintnumbers,
 amsmath,amssymb,
 aps,
pra,
superscriptaddress]{revtex4-1}

\usepackage{graphicx}
\usepackage{dcolumn}
\usepackage{bm}
\usepackage{color,ulem}

\newtheorem{dfn}{Definition}[section]
\newtheorem{thm}[dfn]{Theorem}
\newtheorem{lem}[dfn]{Lemma}

\newcommand{\qed}{\hfill \hbox{\rule[-2pt]{3pt}{6pt}} \par}

\newcommand{\Vp}{V_1^{\prec}}
\newcommand{\CZ}{{\rm CZ}}
\newcommand{\CX}{{\rm CX}}
\newcommand{\tCP}{\widetilde{{\rm CZ}}}
\newcommand{\tCN}{\widetilde{{\rm CX}}}
\newcommand{\tCK}{\widetilde{{\rm CK}}}
\newcommand{\tK}{\widetilde{{\rm K}}}
\newcommand{\tE}{\widetilde{{\rm E}}}
\newcommand{\tX}{\widetilde{{\rm X}}}
\newcommand{\tZ}{\widetilde{{\rm Z}}}
\newcommand{\bra}[1]{\langle {#1} |}
\newcommand{\ket}[1]{| {#1} \rangle}
\newcommand{\X}{{\rm X}}
\newcommand{\Y}{{\rm Y}}
\newcommand{\Z}{{\rm Z}}
\newcommand{\I}{\mathbb{I}}
\newcommand{\U}{{\rm U}}
\newcommand{\acaus}{{\rm ACCZ}}
\newcommand{\J}{{\rm J}}
\newcommand{\Pau}{{\rm P}}

\begin{document}


\title{An analysis of the trade-off between spatial and temporal resources for measurement-based quantum computation}

\author{Jisho Miyazaki}
\affiliation{Department of Physics, Graduate School of Science,\\
The University of Tokyo, 7-3-1 Hongo, Bunkyo-ku, Tokyo, Japan}
\author{Michal Hajdu\v{s}ek}%
\affiliation{Department of Physics, Graduate School of Science,\\
The University of Tokyo, 7-3-1 Hongo, Bunkyo-ku, Tokyo, Japan}%
\affiliation{Singapore University of Technology and Design, 20 Dover Drive, Singapore}
\author{Mio Murao}
\affiliation{Department of Physics, Graduate School of Science,\\
The University of Tokyo, 7-3-1 Hongo, Bunkyo-ku, Tokyo, Japan}%
\affiliation{Institute for Nano Quantum Information Electronics,\\
The University of Tokyo, 4-6-1 Komaba, Meguro-ku, Tokyo}


%


\date{\today}

\begin{abstract}
In measurement-based quantum computation (MBQC),  elementary quantum operations can be more parallelized than the quantum circuit model by employing a larger Hilbert space of graph states used as the resource.  Thus MBQC can be regarded as a method of quantum computation where the temporal resource described by the depth of quantum operations can be reduced compared to the quantum circuit model by using the extra spatial resource described by graph states.  To analyze the trade-off relationship of the spatial and temporal resources,  we consider a method to obtain quantum circuit decompositions of general unitary transformations represented by MBQC on graph states with a certain underlying geometry called generalized flow.   We present a method to translate any MBQC with generalized flow into quantum circuits without extra spatial resource.   We also show an explicit way to unravel acausal gates that appear in the quantum circuit decomposition derived by a translation method presented in [V. Danos and E. Kashefi, Phys. Rev. A {\bf 74}, 052310 (2006)] and that represent an effect of the reduction of the temporal resource in MBQC.  Finally, by considering a way to deterministically simulate these acausal gates, we investigate a general framework to analyze the trade-off between the spacial and temporal resources for quantum computation.

\end{abstract}

\pacs{03.67.Lx, 03.67.Ac, 03.65.Ta}
\maketitle

\section{Introduction}
Measurement-based quantum computation (MBQC) originally proposed in \cite{MBQCPRL} is a framework for quantum computation in which unitary transformations are implemented by measuring qubits of multipartite entangled states. The multipartite entangled states used as resources in MBQC are characterized by graphs specifying to which pairs of qubits entangling operations have been performed to prepare the state and are called graph states~\cite{graph-states-PRA, graph-states-Review}.  The total number of qubits in the graph state is larger than the number of qubits to which unitary transformations are applied.  The graph state extends the ``work-space'' for quantum computation, although the action on the work-space is limited to single-qubit operations (measurements).   Thus they can be regarded as a {\it spatial} resource for quantum computation.  

In MBQC, the choice of a graph state and measurements, which is referred to as a {\it measurement pattern}, specifies the implemented unitary transformation. The choice of measurements depends on the outcomes of previous measurements in order to counter nondeterministic state transformations caused by these measurements.  Such measurements are called as feed-forward measurements. The temporal order of measurements should be carefully chosen to guarantee deterministic implementation of unitary transformations.  The quantum depth of a measurement pattern is determined as the minimum number of steps required for preparing a graph state and for performing the feed-forward measurements when any measurements that are not temporally ordered can be performed in a single step. Quantum depth of a quantum circuit which does not include classically controlled operations depending on measurement outcomes is defined as the number of elementary gates included in the longest dependent sequence of gates in that circuit. Thus the depth can be regarded as a {\it temporal} resource for quantum computation.

For several algorithms including the approximate quantum Fourier transformation \cite{QFT}, it has been shown that MBQC requires smaller quantum depth than a variation of the quantum circuit model without classically controlled operations depending on measurement outcomes \cite{complexity-of-MBQC, Fanout}. This advantage of MBQC originates from the constant-time implementability of any sequence of Clifford gates \cite{MBQCPRA, graph-change-by-measurements} due to the extended work-space by using ancillary qubits of graph states and the feed-forward measurements.   In this case, we can see that the spatial resource (ancillary qubits) is used for reducing the temporal resource (the quantum depth).

Flow \cite{flow} and generalized flow (or {\it gflow} for short) \cite{gflow} are ordering relations on a graph that guarantee the existence of a proper ordering of the measurements required for a deterministic implementation of a unitary transformation by a measurement pattern irrespective of the choices of measurement angles. If flow or gflow exists on a graph, it determines the ordering of measurements, and thus gives an upper bound for the quantum depth of measurement patterns on the graph. These upper bounds are called the depth of flow and gflow, respectively. The depth of gflow on a graph is lower than the depth of flow on the same graph, since gflow is a generalization of flow. There are graphs which have gflow but do not have flow for the same reason.

Given a quantum circuit decomposition of a unitary transformation, we can construct a measurement pattern with depth of flow equal or less than the quantum depth of the original circuit \cite{flow}.	This implies that the depth of flow, and so the depth of gflow, already takes into account the constant time implementability of Clifford gates.

In order to study how the depth of flow and gflow are related to the constant time implementability of Clifford gates, it would be helpful to construct a method to write a circuit decomposition with no ancillary qubits representing the same unitary transformation implemented by a measurement pattern with flow or gflow. Since this compact circuit decomposition does not utilize extra work-space, it includes sequences of Clifford gates that contribute to  the increase of quantum depth of the circuit but not to the depth of flow and gflow.  Thus translations of a unitary represented by MBQC to that of a quantum circuit provide a clue for understanding the trade-off relation between the spacial and temporal resources.

To date, there are three methods to translate a measurement pattern into a compact quantum circuit proposed by \cite{flow, category, compact}.	In \cite{flow}, a translation method called the star pattern transformation (SPT) applicable to measurement patterns on the graph states with flow is presented. If we ignore the depth for implementing Clifford gates, the depth of the resulting circuit coincides with the depth of flow of the original measurement pattern. If we use the SPT to convert a measurement pattern on a graph with gflow but without flow into a quantum circuit, the translation fails and we cannot avoid obtaining an {\it acausal} circuit with ill-defined two qubit gates simultaneously acting in two different steps of time. In \cite{category} and \cite{compact} the authors investigated translation methods applicable also for graphs with gflow but no flow.	The method proposed in \cite{category} based on category theory translates any measurement pattern with gflow into compact circuits and is applicable to a more general class of measurement patterns with no gflow.	The depth of resulting circuits, however, is not analyzed and does not necessarily coincide with the depth of gflow even if the depth for implementing sequences of Clifford gates are assumed to be constant on the circuit.

If acausal gates are allowed to be used in the quantum circuit model, its computational power can be greatly enhanced \cite{PCTCcompt, DCTCcompt}. Authors of \cite{gflow} suggest that the acausal circuits obtained by applying the SPT for measurement patterns of MBQC may efficiently implement the unitary transformation represented by the original measurement pattern.	We further expect that, from a viewpoint of the trade-off between spatial and temporal resources of computation, a measurement pattern with gflow reduces the quantum depth by deterministically simulating acausal gates by utilizing the extra work-space.

In this paper, we propose a new method to translate a measurement pattern on a graph state with gflow into a compact quantum circuit by generalizing the SPT. Based on the new translation method, we clarify the relation between the depth of gflow and constant time implementability of Clifford gates.	We investigate the properties of graphs with gflow and the entanglement structure of the graph states defined by these graphs, and construct the translation method. We show the existence of path covers on graphs with gflow, which was previously shown only for graphs with flow \cite{vicious-circuit}. Local unitary transformations on certain sets of qubits are used for simplifying the entanglement structure of the graph state.

We also show a relation between the circuit decomposition obtained by our method and the acausal circuits obtained by directly applying the SPT for measurement patterns on graph states with gflow but no flow. An operation represented by an acausal two-qubit gate simultaneously acting on two different time positions of the acausal circuit is defined to be consistent with a unitary transformation implemented by the measurement pattern. Finally we discuss how MBQC compresses the quantum depth in connection with the acausal circuit representation.

This paper is organized as follows.
In Sec.~\ref{sec:Preliminaries}, we review MBQC and the properties of a graph corresponding to a graph state used as a resource for deterministic MBQC.	We also reformulate the SPT on graphs with flow.
In Sec.~\ref{sec:pathcover}, we show several graph-theoretical properties of gflow.
In Sec.~\ref{sec:translation}, we present the translation method from a measurement pattern to a quantum circuit using a transformation of a graph with gflow to a graph with flow.
In Sec.~\ref{sec:parallelizing}, the quantum circuit obtained by the method in the previous section is further transformed to parallelize non-Clifford gates.
In Sec.~\ref{sec:SPTforGflow}, we introduce the SPT for graphs with gflow and formally define an acausal circuit.
In Sec.~\ref{sec:acausal_circuit}, we present another translation method from a measurement pattern with gflow to a quantum circuit via an acausal circuit.
In Sec.~\ref{sec:depth}, we discuss the relation between the acausal circuit and the compression of quantum depth.

\section{Preliminaries}\label{sec:Preliminaries}

\subsection{Graph and graph states}
For a given graph $G=(V,E)$ with the vertex set $V$ and the edge set $E \subset \{ \{ u,v \}|u,v \in V,~u \neq v  \}$,
we choose a set of input vertices $I \subset V$ and a set of output vertices $O \subset V$, corresponding to the qubits used to encode the input state and decode the output state respectively.
The triplet $(G,I,O)$ is called an open graph.

If $V'$ is a subset of $V$, $V'^C$ represents the complement of $V'$. We say vertices $u$ and $v$ are connected if $\{ u,v \} \in E$ and denote it by $u \sim v$ ($u \sim_G v$, for specifying a graph $G$).	A neighborhood of vertex $v$ on $G$ is a set of vertices that are connected to $v$ on $G$, and is denoted by $N_G(v)$.	$Odd_G(V_0)$ ($Even_G(V_0)$) represents the odd (even) neighborhood of $V_0 \subset V$ on $G$, i.e. the set of vertices that are connected to the odd (even) number of vertices in $V_0$.
For example, $Odd_G(\{u \})$ is just the neighborhood of vertex $u$.
$V_0 \oplus V_1$ represents the symmetric difference between vertex sets $V_0$ and $V_1$ defined by $V_0 \oplus V_1 = (V_0 \cup V_1) \backslash (V_0 \cap V_1)$.
Note that $Odd (V_0 \oplus V_1)= Odd(V_0) \oplus Odd(V_1)$ \cite{Diestel}.
Vertex sets $V_0$ and $V_1$ are said to be linearly independent in a vertex set $V_2$, when $(V_0 \oplus V_1) \cap V_2 \neq \emptyset$.
Similarly, a set of vertex sets $ \{ V_n { \} }_{n \in \Lambda }$ is said to be a basis of $V'$ if $|\Lambda| = |V'|$ and for any subset $\Gamma \subset \Lambda$, $\bigoplus_{n \in \Gamma} V_n \cap V' \neq \emptyset$.
These relations are understood as a linear independence in the vertex space, where the symmetric difference corresponds to ${\mathbb F}_2$ addition \cite{Diestel}.

The Pauli matrices are denoted by capital letters $\X,\Y$ and $\Z$ in this paper.
The eigenstates of $\Z$ (the computational basis) corresponding to eigenvalues $1$ and $-1$ are represented by $|0 \rangle$ and $|1 \rangle$, respectively.
The eigenstates of $\X$ corresponding to eigenvalues $1$ and $-1$ are represented by $|+ \rangle$ and $|- \rangle$, respectively.  
We define a general controlled-unitary transformation on a set of qubits specified by a set of indices $\mathcal{S}^\prime$ where a qubit specified by index $v \in \mathcal{S}^\prime $ is a controlled qubit and 
a unitary transformation $\U$ is applied on the rest of qubits specified by a set $\mathcal{S}:= \mathcal{S}^\prime  \backslash v$ only when the controlled qubit is in $\ket{1}$, otherwise no transformation is applied, namely,
\begin{eqnarray}
	\textrm{CU}_{v;\mathcal{S}} := |0\rangle \langle 0|_v \otimes \I + |1\rangle \langle 1|_v \otimes \U.
	\label{eq:controlled-U}
\end{eqnarray}
If $\mathcal{S} = \{ u \}$, $\textrm{CU}_{v;\mathcal{S}}$ is also represented by $\textrm{CU}_{v;u}$.	In particular, $\CZ_{v;u}$ and $\CX_{v;u}$ are called a $\CZ$-gate and a CNOT-gate, respectively.

A quantum state corresponding to an open graph is called an open graph state and is constructed in the following way.
First we prepare a qubit system on each vertex of the graph $G$.
Each qubit is labeled by the index of the corresponding vertex.
All qubits with indices in $I^C$ are prepared in the $|+ \rangle$ state whereas the qubits with the indices in $I$ are prepared in a joint input state $\ket{\phi}$.  
Next, $\CZ$-gates are applied to all pairs of qubits corresponding to adjacent vertices, namely, the qubits with indices connected by edges $E$ of $G$.
We denote an unitary transformation $\U$ acting on qubits of the graph states by $\widetilde{\U}$ in order to distinguish it from a unitary transformation acting on logical qubits of the corresponding circuit.
Then an open graph state $\ket{G}_\phi$ of an open graph $G$ with an input state $\ket{\phi}_I$ is represented by 
\begin{eqnarray}
\nonumber	| G \rangle_{\phi} = \tE_G  | \phi \rangle_I |+ \rangle_{I^C},
\end{eqnarray}
where
\begin{eqnarray}
\nonumber	\tE_G = \prod_{\{u,v\} \in E} \tCP_{u;v}
\end{eqnarray}
This state is stabilized by $\tK_v := \tX_v \tZ_{N(v)}$($v \in I^C$), namely, $\tK_v | G \rangle_{\phi} = | G \rangle_{\phi}$.

\subsection{Flow}
After preparing the open graph state, the unitary transformation is implemented by performing projective measurements on each qubit in $O^{C}$.
The measurement operators are defined by $\{ | \pm_{\alpha_v} \rangle \langle \pm_{\alpha_v} | \}$ where
\begin{eqnarray}
\nonumber	| \pm_{\alpha_v} \rangle := \left( |0 \rangle \pm e^{i \alpha_v}|1 \rangle \right) / \sqrt{2},
\end{eqnarray}
and $\alpha_v \in [0, 2\pi)$ represents the measurement angle depending on the vertex $v \in O^C$.
If we obtain a measurement result ``$-$'' on a qubit, we adjust the measurement angles of future measurements so that the quantum computation proceeds as if we had obtained the result ``$+$''.
Therefore the unitary transformation implemented by deterministic MBQC on an open graph $(G,I,O)$ is proportional to
\begin{eqnarray}
	\bigotimes_{u \in O^C} \langle +_{\alpha_u}| \tE_G |+\rangle_{I^C}.
	\label{eq:wholeunitary}
\end{eqnarray}
The dependency relation of the measurement angles determines the ordering of measurements.
Flow \cite{flow} is an ordering relation guaranteeing deterministic computation, and is a pair $(f, \prec)$ of a function $f:O^C \rightarrow I^C$ and a partial order $\prec$ satisfying the following conditions
\begin{description}
	\item[f-1] $u \prec f(u)$
	\item[f-2] $u \in N(f(u))$
	\item[f-3] $\forall v \in N(f(u))$, $u = v$ or $u \prec v$.
\end{description}
Graph theoretical properties of flow are analyzed in \cite{vicious-circuit}.
A path cover is an important property of a graph with flow for understanding the correspondence with the circuit model.
It is defined by the following.
\begin{dfn}[path cover\cite{vicious-circuit}]\label{def:path_cover}
	Let $(G, I,O)$ be an open graph.
	A collection $P_f$ of (possibly trivial) directed paths in $G$ is a path cover of $(G, I,O)$ if
	\begin{itemize}
		\item each $v \in V (G)$ is contained in exactly one path (i.e. the paths cover $G$ and they are vertex-disjoint);
		\item each path in $P_f$ is either disjoint from $I$, or intersects $I$ only at its initial point;
		\item each path in $P_f$ intersects $O$ only at its final point.
	\end{itemize}
	\end{dfn}
For any open graph with flow, a unique path cover $P_f$ is defined by
\begin{eqnarray}
\nonumber	v_0 \rightarrow v_1 \rightarrow ... \rightarrow v_n \in P_f \Leftrightarrow v_n \in O \wedge f(v_i)=v_{i+1}~(\forall i).\\
\label{eq:pathcover}
\end{eqnarray}
Note that this definition of a path cover is more restrictive compare to the notion commonly used in graph theory where a path cover is a set of disjoint paths on a directed graph and does not necessarily connect vertices in $I$ and $O$ \cite{Diestel}.

\subsection{Circuits and measurement patterns with flow}\label{cf}
A {\it measurement pattern} consists of the open graph $(G,I,O)$, the ordering of measurements and the measurement angles $\alpha_v (v \in O^C)$ which may depend on the outcomes of previous measurements.	The measurement pattern determines how to prepare the qubits, how to entangle them and how to perform measurements.

{\it Star pattern transformation} (SPT) is a method \cite{flow} (see also Ref.~\cite{flow-to-circuit} in this context) to translate a unitary transformation implemented by a measurement pattern with flow to a circuit decomposition,
in such a way that each measurement in the measurement pattern corresponds to an elementary gate in the circuit.
In this subsection, we reformulate this method to be easily extendable for measurement patterns with gflow.

The procedure of the SPT is divided into three parts.

\begin{figure}
	\centering
	\includegraphics[width=8cm]{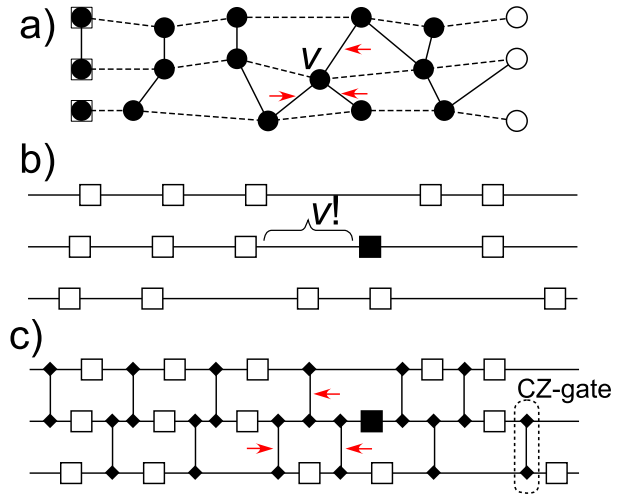}
	\caption{(color online) (a) An open graph with flow. Circles and lines represent the vertices and edges, respectively. The boxed vertices represent inputs and the white vertices are outputs by following the notation of MBQC presented in \cite{flow}. The set of dashed edges represents the path cover and the vertex incident with the edges pointed by arrows is labeled by $v$.	(b) A quantum circuit corresponding to the open graph given by (a) obtained by process (ii) of SPT (Sec.~\ref{cf}). The three wires correspond to the path cover, and boxes represents the $\J$-gates. The black box represents a particular $\J$-gate $\J(\alpha_v)$ assigned for the vertex labeled by $v$. The position $v!$ denotes a region on the wire between $\J(\alpha_v)$ and $\J(\alpha_{f^{-1}(v)})$.	(c) A quantum circuit representing the measurement pattern on the open graph given by (a). The $\CZ$-gates pointed by arrows correspond to the edges pointed by arrows on the graph.}
	\label{fig:SPT_flow}
\end{figure}

\begin{description}
	\item[($i$)] We regard each path in $P_f$ as a wire that represents a Hilbert space of a qubit ${\mathbb C}^2$ in the circuit.
\end{description}
The wire in the circuit corresponding to the path including vertex $v$ on the graph is also labeled by $v$.
A wire labeled by a flow image $f(v)$ of a vertex $v$ is identical to the wire labeled by $v$.
\begin{description}
	\item[($ii$)] We place a $\J$-gate $\J(\alpha_v)$ defined by
\begin{eqnarray}
\nonumber	\J(\alpha_v) := \frac{1}{\sqrt{2}} \left( \begin{array}{cc}
			1 & e^{- i \alpha_v} \\
			1 & - e^{- i \alpha_v} \\
			\end{array}\right)
\end{eqnarray}
on wire $v$ if qubit $v$ is measured at an angle $\alpha_v$.
\end{description}
These $\J$-gates must be placed so that $\J(\alpha_u)$ acts before $\J(\alpha_v)$ if $u \prec v$ on the graph.
We sometimes have to specify not only wires but also the position in the wire on which a gate acts.
This position indicates the timing when the gate acts on the qubit represented by the wire.
The position between $\J(\alpha_{f^{-1}(v)})$ and $\J(\alpha_{v})$ in the wire $v$ is labeled by $v!$.
If $v$ is a starting vertex of a path cover, the label $v!$ represents the position before the gate $\J(\alpha_v)$ in the wire $v$, (see FIG~\ref{fig:SPT_flow}.)
\begin{description}
	\item[($iii$)] If there is an edge between vertices $u$ and $v$, we place $\CZ_{u!;v!}$.
			The ordering of multiple $\CZ$-gates corresponding to non-path edges incident from the same vertex is determined by the partial ordering of the vertices on the other side of the edges corresponding to the $\CZ$-gates.
\end{description}
We define a binary relation $\prec_p$ on the positions in the circuit by
\begin{eqnarray}
	u! \prec_p v! \Leftrightarrow && \exists p\in P_f~ s.t.~ (u \rightarrow v) \in p \\
	\label{eq:gateseq1}
\nonumber					&\vee& \exists p\in P_f , \exists w \in N(v) \backslash \{ u \}~ s.t. (u \rightarrow w) \in p\\
	\label{eq:gateseq2}
\end{eqnarray}
This binary relation can be defined on any circuit corresponding to a measurement pattern on a graph with a path cover.
The relation $\prec_p$ must be a partial order to have a consistent gate sequence.
If the path cover $P_f$ is defined by flow according to Eq.(\ref{eq:pathcover}), $u! \prec_p v! \Leftrightarrow u \prec v$ holds.
There is a consistent gate sequence on any circuit corresponding to a measurement pattern on a graph with flow because $\prec$ is a partial order.
Conversely, if an open graph with a path cover does not have flow, the binary relation $\prec_p$ on the corresponding circuit is not a partial order, and the gate sequence is not well-defined as a quantum circuit.

By reversing this method, we obtain a measurement pattern representation of a unitary transformation from a circuit representation whose elementary gates are given by $\J$-gates and $\CZ$-gates \cite{flow-to-circuit}.

\subsection{gflow}
\begin{figure}
	\centering
	\includegraphics[width=4cm]{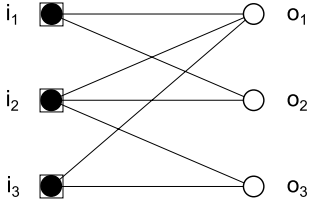}
	\caption{An open graph with gflow but no flow. The input and output vertices are $I = \{ i_1, i_2, i_3 \}$ and $O = \{ o_1, o_2 , o_3 \}$, respectively.
		 The last layer $V_0^{\prec}$ with regard to the maximally delayed gflow is $O$, and the second last layer $\Vp$ is $I$.
		 The values of the maximally delayed gflow function $g$ are following: $g(i_1) = \{ o_1 , o_3 \}$, $g(i_2) = \{ o_1 , o_2 , o_3 \}$, $g(o_3) = \{ o_1 , o_2 \}$.
		 The modified gflow $(g_V,\prec_V )$ that satisfies Eqs.~(\ref{eq:gv1}),~(\ref{eq:gv2}) is $g_V(i_1) = g(i_1) = \{ o_1 , o_3 \}$, $g_V(i_2) = g(i_1) \oplus g(i_2) = \{ o_2 \}$, $g_V(i_3) = g(i_2) \oplus g(i_3) = \{ o_3 \}$.}
	\label{fig:gYfN}
\end{figure}

Gflow is defined as follows
\begin{dfn}[\cite{gflow} definition.3]\label{gflow}
	Let $(G,I,O)$ be an open graph.
	Let $g:O^c \rightarrow 2^{I^c}$ be a function on non-output vertices to the power set of non-input vertices, and $\prec$ be a strict partial order on vertices.
	The pair $(g, \prec)$ is a gflow of the open graph if it satisfies the following three conditions
	\begin{description}
		\item[g-1] $\forall v \in g(u)$, $u \prec v$
		\item[g-2] $u \in Odd(g(u))$
		\item[g-3] $\forall v \in Odd(g(u))$, $u = v$ or $u \prec v$.
	\end{description}
\end{dfn}
Gflow is a generalization of flow in the sense that $g$ can take a set of vertices whereas the flow function $f$ can take only one vertex.
There are graphs that have gflow but do not have flow.
In this case, the SPT does not lead a well defined circuit.
For example, an open graph presented in FIG.~\ref{fig:gYfN} has a path cover, so we can define wires of the circuit for this graph.
However we cannot assign all $\CZ$-gates in a way obeying a well-defined ordering of gates (FIG.~\ref{fig:gYfN_acausal}(a)).

The strict partial order of gflow induces a temporal ordering that the sequence of measurements and corrections must follow.
Vertices that do not have ordering between each other are said to be in the same layer.
\begin{dfn}[Layers\cite{max-delayed-gflow}]\label{layer}
	Let (G,I,O) be an open graph with gflow (g,$\prec$). Layers $V_k^{\prec}$ of this gflow are defined as
	\begin{eqnarray}
	\nonumber	V_0^{\prec} \equiv& {\max}_{\prec} V(G) \\
	\nonumber	V_k^{\prec} \equiv& {\max}_{\prec} V(G) \backslash {\cup}_{i<k} V_i^{\prec} \hspace{8pt} (k>0),
	\end{eqnarray}
	where the maximization in terms of the relation $\prec$ is defined by ${\max}_{\prec} X \equiv \{ u \in X| \forall v \in X, \lnot (u \prec v) \}$.
	\end{dfn}
When there are $d_{\prec} +1$ layers $V_0^{\prec}, ..., V_{d_{\prec}}^{\prec}$, $d_{\prec}$ is called {\it depth of the gflow}.
Measurements of qubits corresponding to the vertices belonging to the same layer can be performed simultaneously.
The depth of the gflow represents the number of rounds of simultaneous measurements required according to the gflow.

\begin{dfn}[delay]\label{delay}
	A gflow $(g,\prec )$ is more delayed than $(g',\prec ')$ if and only if
	\begin{eqnarray}
		\forall k, \hspace{8pt} |{\cup}_{i=0}^k V_i^{\prec} | \geq |{\cup}_{i=0}^k V_i^{\prec '} |
	\label{eq:delay}
	\end{eqnarray}
	and there is a number specified by $k$ with which the inequality (\ref{eq:delay}) becomes strict.
	\end{dfn}
	
In general, gflow is not unique and so is the depth of gflow.
The gflow with minimal depth on an open graph is called maximally delayed gflow.

Maximally delayed gflow has the following properties
\begin{eqnarray}
\nonumber	V_0^{\prec} = O, \\
\nonumber	V_1^{\prec} = \{ v \in O^C| \exists S_v \subset O, Odd(S_v) \cap O^C = \{ v \} \}.
\end{eqnarray}
These properties are used extensively in our analysis.

\section{Path covers for gflow}\label{sec:pathcover}
\begin{figure}
	\centering
	\includegraphics[width=4cm]{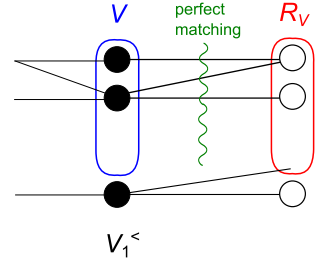}
	\caption{(color online) Perfect matching between $V \subset V_1^{\prec}$ and $R_V \subset O$ (Lemma~\ref{matching}).}
	\label{fig:matching}
\end{figure}

In this section we show the existence of path cover on graphs with gflow.
Paths of the path cover will be regarded as the wires in the circuit decomposition similarly to the cases of graphs with flow presented in Sec.~\ref{cf}.
We prove the existence of the path cover by construction. The first step is to find a matching between the output and the penultimate layer.
Following lemmas, which are similar to the lemmas presented in Section 2.3.2 of Ref.~\cite{global-circuit-optimization}, are used for the proof.
\begin{lem}\label{matching}
	Let $(G,I,O)$ be an open graph with gflow, with its layers $\{ V_i^{\prec} { \} }_{i=0,...,d} (V_0^{\prec} =O)$ defined by maximally delayed gflow $(g,\prec)$.
	For all subsets $V \subset {\Vp}$, there is a subset $R_V \subset O$ and a gflow $(g_V, \prec_V)$ satisfying the following four conditions,
	\begin{description}
	\item[R-a] $|R_V| = |V|$,
	\item[R-b] $\{ R_V \cap g(v) { \} }_{v \in V} $ becomes a basis of $R_V$,
	\item[R-c] There is a perfect matching between $R_V$ and $V$, and the edges of the matching are chosen from real gflow edges of $(g_V, \prec_V)$, where a real gflow edge refers to the edge $(x, y) \in E$ satisfying $y \in g(x)$ (FIG.\ref{fig:matching}).
	\item[R-d] $Odd(g_V(v)) \cap (\Vp \backslash V) = \emptyset$ $(\forall v \in V)$.
\end{description}
\end{lem}

$Proof$)
The proof is by induction with respect to $|V|$.
The statement holds for the case of $|V|=|\{ v \}|=1$, if we choose $(g_V, \prec_V) = (g,\prec)$ and $R_V$ to be any vertex $\{ r_1 \}$ in $g(v)$.
We assume that there exists a subset $R_V$ that satisfies conditions {\bf R-a} to {\bf R-d} for any $V$ with $|V| \leq m$.
Let $V_{n} = \{ v_1 , v_2 , ... , v_{n} \}$ ($\forall n \leq m+1$).
From the assumption, there is a subset $R_{V_m} \subset O$ that satisfies conditions {\bf R-a} to {\bf R-d}.
We denote the new gflow described in {\bf R-c} and {\bf R-d} by $(g_{V_m}, {\prec}_{V_m})$.
From {\bf R-b}, there exists a vertex set $U_{m+1} \subset V_m$ such that
\begin{eqnarray}
	R_{V_m} \cap g(v_{m+1}) = R_{V_m} \cap \bigoplus_{v \in U_{m+1}} g(v).%
\label{eq:Um}
\end{eqnarray}
Let us define a subset $O_{m+1}$ of $O$ by
\begin{eqnarray}
	O_{m+1} := g(v_{m+1}) \oplus \bigoplus_{v \in U_{m+1}} g(v).
\label{eq:Om}
\end{eqnarray}
By definition, $O_{m+1} \cap R_{V_m} = \emptyset$.
We define a function $g_{V_{m+1}}:O^C \rightarrow 2^{I^C}$ by
\begin{eqnarray}
	g_{V_{m+1}}(v) := \left \{ \begin{array}{rcl}
			O_{m+1} & \mbox{for} & v = v_{m+1} \\
			g_{V_m}(v) & \mbox{otherwise.} & \\
			\end{array}\right.
\label{eq:def_gV}
\end{eqnarray}
For any $n \leq m+1$,
\begin{eqnarray}
\nonumber	Odd(g_{V_{m+1}}(v_n)) \backslash O &=& Odd(O_n) \backslash O \\
\nonumber				&=& (Odd(g(v_n)) \backslash O) \oplus \bigoplus_{v \in U_n} Odd(g(v)) \backslash O\\
\nonumber				&=& \{ v_n \} \oplus \bigoplus_{v \in U_n} \{ v \} \\
\nonumber				&=& \{ v_n \} \cup U_n
\end{eqnarray}
holds.
For the case $n=m+1$, $v_{m+1} \in Odd(g_{V_{m+1}}(v_{m+1}))$ implies the existence of odd number of edges between $v_{m+1}$ and $g_{V_{m+1}}(v_{m+1})$.
We choose a vertex $r_{m+1}(\sim v_{m+1})$ from $g_{V_{m+1}}(v_{m+1})$ for the matching with $v_{m+1}$.
This vertex is not included in $R_{V_{m}}$ because $O_{m+1} \cap R_{V_m} = \emptyset$.
For later use, we define a function $h$ on the vertices inductively by
\begin{eqnarray}
	h(v_{m+1}) = r_{m+1}.
\label{eq:successor}
\end{eqnarray}
The domain of this function becomes $O^C$ after the induction is finished.

Now we have to define a partial ordering $\prec_{V_{m+1}}$ for function $g_{V_{m+1}}$.
Because $Odd(g_{V_{m+1}}(v_{m+1})) \backslash O = v_{m+1} \cup U_{m+1}$, $v_{m+1} \prec_{V_{m+1}} U_{m+1}$ must hold.
This is allowed if $v_{m+1} \notin Odd(g_{V_{m}}(v))$ for any vertex $v \in U_{m+1}$, which is guaranteed by the assumption {\bf R-d}.
Thus $g_{V_{m+1}}$ and the ordering inductively defined by
\begin{eqnarray}
\nonumber	v \prec_{V_{m+1}} u \Leftrightarrow (v=v_{m+1} \wedge u \in U_{m+1} \cup O) \vee (v \prec_{V_m} u) \\
\label{eq:partial_gv}
\end{eqnarray}
is a gflow on the open graph $(G,I,O)$.

Now we define
\begin{eqnarray}
	R_{V_{m+1}} := R_{V_m} \cup \{ r_{m+1} \}.
	\label{eq:defRm+1}
\end{eqnarray}
The gflow $(g_{V_{m+1}},\prec_{V_{m+1}})$ and $R_{V_{m+1}}$ satisfy {\bf R-c} because $v_{m+1}$ is connected to $r_{m+1} \in O_{m+1} \backslash R_{V_m} \subset O \backslash R_{V_m}$ and $r_{m+1} \in O_{m+1} = g_{V_{m+1}}(v_{m+1})$.
The condition {\bf R-d} follows from the definition of $(g_{V_{m+1}},\prec_{V_{m+1}})$.

It remains to show {\bf R-b} since {\bf R-a} is trivial.
By assumption, $\{ R_{V_m} \cap g(v) { \} }_{v \in V_m}$ is the basis of $R_{V_m}$.
This implies that a union of $\{ R_{V_{m+1}} \cap g(v) { \} }_{i \in V_m}$ and $\{ r_{m+1} \} = R_{V_{m+1}} \cap g_{V_{m+1}}(v_{m+1})$ forms the basis of $R_{V_{m+1}}$.
Then from the definition of $g_{V_{m+1}}(v_{m+1})$, $\{ R_{V_{m+1}} \cap g(v) { \} }_{v \in V_{m+1}}$ becomes the basis of $R_{V_{m+1}}$.

Thus the statement of the lemma holds for $|V|=m+1$, which concludes the proof.
\qed

We give a particular name to the gflow constructed by this lemma for convenience.
\begin{dfn}[matching gflow]\label{dfn:matching_gflow}
	If $V= \Vp = \{ v_1, v_2,...,v_l \}$ ($l = |\Vp|$), we call the gflow $(g_{V_l},\prec_{V_l})$ constructed inductively by Eqs.~(\ref{eq:Um}),(\ref{eq:Om}),(\ref{eq:def_gV}),(\ref{eq:defRm+1}) and (\ref{eq:partial_gv})  {\rm a matching gflow} of $(G,I,O)$, and denote it by $(g_V,\prec_V)$.
	We call the function $h:\Vp \rightarrow O$ defined by Eq.~(\ref{eq:successor}) {\rm a successor function} of $(g_V,\prec_V)$.
\end{dfn}

Lemma~\ref{matching} guarantees the existence of a matching between $\Vp$ and a suitable subset $R_V$ of $O$.
The following lemma helps to find the matching between other layers by reducing $R_V$.
\begin{lem}\label{removeOUT}
	Let $(G,I,O)$ be an open graph with gflow, with its layers $\{ V_i^{\prec} { \} }_{i=0,...,d} (V_0^{\prec} =O)$ defined by maximally delayed gflow $(g,\prec)$.
	If the subset $R \subset O $ satisfies {\bf R-a} and {\bf R-b} of Lemma~\ref{matching} with $V = {\Vp}$,
	then a new open graph $(G \backslash R,I \backslash R, V_1^{\prec} \cup (O \backslash R))$ has maximally delayed gflow with the same ordering $\prec$.
\end{lem}
\begin{figure}
	\centering
	\includegraphics[width=6cm]{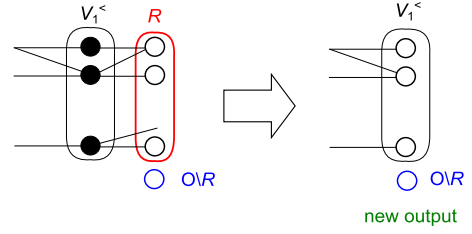}
	\caption{(color online) Removing $R$ from $O$ (Lemma~\ref{removeOUT}).}
	\label{fig:removing}
\end{figure}
$Proof$)
We define a reduced open graph $(G', I', O') = (G \backslash R, I \backslash R, V_1^{\prec} \cup (O \backslash R))$, and construct a gflow on this open graph.
Because $\{ R \cap g(v) { \} }_{v \in {\Vp}} $ is a basis of $R$, for all $v \in V(G') \backslash O'$, there is a subset $V_v \subset {\Vp}$ such that
\begin{eqnarray}
\nonumber	g(v) \cap R = \bigoplus_{u \in V_v} g(u) \cap R.
\end{eqnarray}
We define a new gflow function $g':O^c \rightarrow 2^{I^c}$ by
\begin{eqnarray}
\nonumber	g'(v) := g(v)\backslash R \oplus \bigoplus_{u \in V_v} g(u) \backslash R.
\end{eqnarray}
It can be checked that $v \prec w$ $(\forall w \in g'(v))$ holds for the original partial order of gflow $(g,\prec)$, due to $\oplus_{u \in V_v} g(u) \subset O$.
The following calculation shows that $v \prec w$ $(\forall w \in Odd(g'(v)))$ also holds.
\begin{eqnarray*}
	&& Odd_{G'}(g'(v)) \hspace{3cm} \\
	&=& Odd_G \left( g(v) \backslash R \oplus \bigoplus_{u \in V_v} g(u) \backslash R \right) \backslash R \\
	&=& Odd_G \left( g(v) \backslash R \oplus \bigoplus_{u \in V_v} \left( ( g(u) \cap R ) \oplus g(u) \right) \right) \backslash R \\
	&=& Odd_G \left( g(v) \backslash R \oplus \left( g(u) \cap R \right) \oplus \bigoplus_{u \in V_v} g(u) \right) \backslash R \\
	&=& Odd_G \left( g(v) \oplus \bigoplus_{u \in V_v} g(u) \right) \backslash R \\
	&=& \left[ Odd_G \left( g(v) \right) \oplus Odd_G \left( \bigoplus_{u \in V_v} g(u) \right) \right] \backslash R \\
	&=& v \oplus V_v \oplus (\text{subset of } O) \backslash R \\
	&=& v \oplus (\text{subset of } O').
\end{eqnarray*}

It remains to show that $(g', \prec)$ is maximally delayed in $G'$.
Let $(g_m, {\prec}_m)$ be the maximally delayed gflow of $(G',I',O')$, then
\begin{eqnarray*}
	&& V_1^{{\prec}_m} \\
	&=& \{ v \in G' \backslash O'| ^\exists S_v \subset O', Odd_{G'} (S_v) \backslash O' = \{ v \} \} \\
	&=& \{ v \in G \backslash (O \cup {\Vp}) | \\
		&& ^\exists S_v \subset ( O \cup {\Vp} ) \backslash R, Odd_{G} (S_v) \backslash ( O \cup {\Vp} ) = \{ v \} \} \\
	& \subset & \{ v \in G \backslash (O \cup {\Vp}) | \\
		&& ^\exists S_v \subset ( O \cup {\Vp} ), Odd_{G} (S_v) \backslash ( O \cup {\Vp} ) = \{ v \} \} \\
	&=& V_2^{\prec}.
\end{eqnarray*}
Since $V_2^{\prec}$ is the first layer in $(G',I',O')$ with respect to $(g', \prec)$, $|V_2^{\prec}| \leq |V_1^{{\prec}_m}|$ holds.
Thus we have $V_2^{\prec} = V_1^{{\prec}_m}$, and we can show the equality of all layers (i.e. identity of ${\prec}_m$ and ${\prec}$) by induction.
\qed

If we consecutively apply Lemma~\ref{removeOUT} with a subset to be removed chosen according to Lemma~\ref{matching} (by taking $V = {\Vp}$), the resulting set of edges used for the matchings forms a path cover.
\begin{thm}\label{pathcover}
	If an open graph $(G,I,O)$ has gflow, then there exists a path cover.
	Each edge of the paths can be chosen from real gflow edges of some gflow on $(G,I,O)$.
	\end{thm}
The path cover defined in this way is not necessarily unique.
There is an arbitrariness in choosing the subset $R$ and the matching between $R$ and ${\Vp}$.

We note that Lemma~\ref{removeOUT} solely implies that $|V_k^{\prec}| \leq |O|$ for all $k$, since the number of output vertices is not changed by the use of Lemma~\ref{removeOUT} (i.e. $|O'|=|V_1^{\prec} \cup (O \backslash R)| = |O|$).
This result implies that the depth of gflow is upper bounded by $|V|/|O|$.

\section{Translation from MBQC into quantum circuit}\label{sec:translation}

The path cover we have constructed in Sec.~\ref{sec:pathcover} is used as a wire of the corresponding circuit decomposition.
We divide the unitary transformation represented by a measurement pattern into a step by step unitary transformation implemented between each of the layers $V_{i}^\prec \rightarrow V_{i-1}^\prec$.
For simplicity of notation, we define the following two multi-qubit gates:
\begin{eqnarray}
\nonumber	\CZ_{v;\mathcal{S}} := \prod_{u \in \mathcal{S}} \CZ_{v;u}
\end{eqnarray}
and
\begin{eqnarray}
\nonumber	\CX_{v;\mathcal{S}} := \prod_{u \in \mathcal{S}} \CX_{v;u}.
\end{eqnarray}

\begin{figure}
	\centering
	\includegraphics[width=4cm]{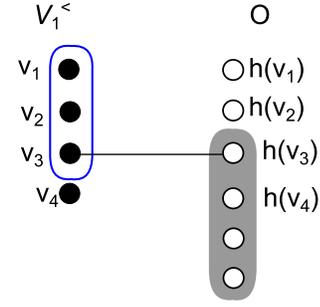}
	\caption{(color online) Properties of the gflow $(g_V ,\prec_V)$. There is an edge between $v_i$ and $h(v_i)$. A gflow image $g_V (v_3)$ is in the shaded region (Eq.~(\ref{eq:gv1})). $Odd_{G'} \left( g_V (v_i) \right)$ is in the circled region (Eq.~(\ref{eq:gv2})).}
	\label{fig:prop_newgflow}
\end{figure}

\begin{lem}\label{lem:gflow_to_flow}
	Let $(G,I,O)$ be an open graph with maximally delayed gflow $(g,\prec)$.
	Let us remove all the edges inside $O$ from $G$ and denote the resulting open graph by $(G',I,O)$, namely,
	\begin{eqnarray}
	\nonumber	\prod_{\{ u,v \} \in E,~ u,v\in O} \tCP_{u;v} | G' \rangle = | G \rangle.
	\end{eqnarray}
	Then the state
	\begin{eqnarray}
	\nonumber	\tCN_{h(v_n); g_V(v_n) \oplus h(v_n)} ... \tCN_{h(v_1); g_V(v_1) \oplus h(v_1)} | G' \rangle
	\end{eqnarray}
	is also an open graph state, where $(g_V,\prec_V)$ is a matching gflow of $(G,I,O)$, and $h$ is the successor function of $(g_V,\prec_V)$.
	This graph does not have any edges between $R_V$ and the outside of $\Vp$.
	The subgraph of this graph consisting of $\Vp$ and $R_V$ has flow if the input set is $\Vp$ and the output set is $R_V$.
	\end{lem}

$Proof$)
By construction, the gflow $(g_V , {\prec}_V)$ has the properties given by
\begin{eqnarray}
	g_V (v_i) \in O \backslash \{ h(v_1), ..., h(v_{i-1}) \},
	\label{eq:gv1}
\end{eqnarray}
\begin{eqnarray}
	Odd_{G'} \left( g_V (v_i) \right) \in \{ v_1, ..., v_i \},
	\label{eq:gv2}
\end{eqnarray}
where $Odd_{G'}$ represents the odd-neighborhood on the graph $G'$.
Let us denote the set $\left( g_V(v_i) \oplus h(v_i) \right) \cup Odd_{G'}  \left( g_V (v_i) \oplus h(v_i) \right)$ by $W(v_i)$.
The open graph state $|G' \rangle$ is a stabilizer state of the operator
\begin{eqnarray}
	\tK(g_V(v_i)) &=& \tX_{g_V(v_i) \oplus h(v_i)} \tZ_{Odd_{G'} \left( g_V (v_i) \oplus h(v_i) \right)} \\
	\label{eq:matching gflow stabilizer}
\nonumber	&=& \tX_{g_V(v_i) \oplus h(v_i)} \tZ_{Odd_{G'} \left( g_V (v_i) \right) \oplus N_{G'} \left( h(v_i) \right)} \\
\nonumber	&=& \tX_{g_V(v_i) \oplus h(v_i)} \tZ_{Odd_{G'} \left( g_V (v_i) \right)} \tZ_{N_{G'} \left( h(v_i) \right)}.
\end{eqnarray}
The stabilizer $\tK(g_V(v_i))$ is a multi-qubit local unitary transformation acting on the vertices in $W(v_i)$.
It follows that the controlled version of the stabilizer $\tCK(g_V(v_i))_{h(v_i);W(v_i)}$ also stabilizes $| G' \rangle$ (The proof is given in Appendix \ref{sec:append_stabi}.
If we apply $\tCK(g_V(v_i))_{h(v_i);W(v_i)}$ on $| G' \rangle$,
\begin{eqnarray}
	| G' \rangle &=& \tCK(g_V(v_i))_{h(v_i);W(v_i)} | G' \rangle \\
	\label{eq:controlled stabilizer}
			&=& \tCN_{h(v_i); g_V(v_i) \oplus h(v_i)} | G'' \rangle,
	\label{eq:change}
\end{eqnarray}
where on the graph $G''$, all edges incident to $h(v_i)$ are removed by $\widetilde{\CZ}_{h(v_i);N_{G'} \left( h(v_i) \right)}$ and new edges are created between $h(v_i)$ and $Odd_{G'} \left( g_V (v_i) \right)$.
The subgraph on which $\tCK(g_V(v_j))$ acts does not include any vertices in $\{ h(v_1), ..., h(v_j) \}$ (Eq.~(\ref{eq:gv1})).
Therefore all $\tCK(g_V(v_j))_{h(v_j);W(v_j)}$ ($\forall j>i$) are stabilizer operators on $|G '' \rangle$.
We define the sequence of open graphs $\{ G_i \}_{i=0,...,n}$ 
inductively by
\begin{eqnarray}
\nonumber	| G_{i-1} \rangle &=& \tCK(g_V(v_i))_{h(v_i);W(v_i)} | G_{i-1} \rangle \\
				&=& \tCN_{h(v_i); g_V(v_i) \oplus h(v_i)} | G_i \rangle ,
	\label{eq:def_sequence_G}
\end{eqnarray}
where $G_0 := G'$.
The open graph state $| G' \rangle$ is now represented as
\begin{eqnarray}
\nonumber	| G' \rangle &=& \tCN_{h(v_1); g_V(v_1) \oplus h(v_1)} | G_1 \rangle \\
\nonumber		&=& \tCN_{h(v_1); g_V(v_1) \oplus h(v_1)} \tCN_{h(v_2); g_V(v_2) \oplus h(v_2)} | G_2 \rangle \\
\nonumber		&=& ... \\
\nonumber		&=& \tCN_{h(v_1); g_V(v_1) \oplus h(v_1)} ... \tCN_{h(v_n); g_V(v_n) \oplus h(v_n)} | G_n \rangle . \\
	\label{eq:sequense_G}
\end{eqnarray}
\begin{figure}
	\centering
	\includegraphics[width=6cm]{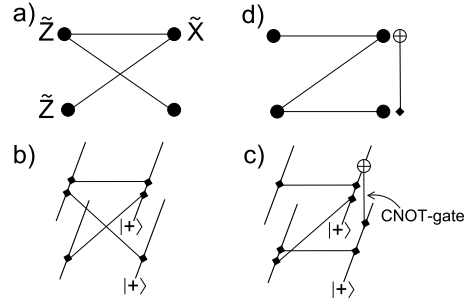}
	\caption{(a) An open graph state. The left two vertices are input qubits and the right two are the outputs. A product of the Pauli $\tX$ and two Pauli $\tZ$ operators described in the figure form a stabilizer of the graph state.
		 (b) The circuit description of the same graph state. The four lines from the bottom left to the top right represent physical qubits on the vertices. (They are not the wires corresponding to the path cover.)
		 (c) The equivalent circuit description including a CNOT-gate.
		 (d) The graph state obtained by applying a CNOT-gate. This state is equivalent to the graph state represented by (a).}
	\label{fig:graph_change}
\end{figure}
Note that the CNOT-gates act inside the output vertices.
The edges incident from $R_V$ to $G_n$ are 
\begin{eqnarray}
\nonumber	h(v_i) \sim v ~ (v \in Odd_{G'} \left( g_V (v_i) \right), \forall i ).
\end{eqnarray}
Since
\begin{eqnarray}
\nonumber	N_{G_n} (h(v_i)) = \{ v \in Odd_{G'} \left( g_V (v_i) \right) \} = Odd_{G'} \left( g_V (v_i) \right),
\end{eqnarray}
$h$ turns out to be a flow from $\Vp$ to $R_V$ with the same partial order to $\prec_V$ on $\Vp \cup R_V$.
\begin{figure}
	\centering
	\includegraphics[width=4cm]{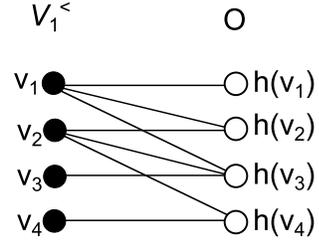}
	\caption{A typical shape of the graph $G_n$. A vertex $h(v_i)$ is connected to vertices $v_j$ where $j \leq i$.}
	\label{fig:graphGn}
\end{figure}
\qed

Now we find the circuit decomposition of the unitary transformation implemented by the last step $\Vp \rightarrow R_V$.
The map represented by Eq.~(\ref{eq:wholeunitary}) and implemented by the measurement pattern on the graph $(G,I,O)$ is
\begin{eqnarray}
\nonumber	&& \bigotimes_{v \in O^C} \langle +_{\alpha_v} |G \rangle \\
	&=& \prod_{\{ u,v \} \in E,~ u,v\in O} \tCP_{u;v} \bigotimes_{v \in O^C} \langle +_{\alpha_v} |G' \rangle \\
	&=& \prod_{\{ u,v \} \in E,~ u,v\in O} \tCP_{u;v} \\
\nonumber	&& \tCN_{r_1; g_V(v_1) \oplus r_1} ... \tCN_{r_n; g_V(v_n) \oplus r_n} \bigotimes_{v \in O^C} \langle +_{\alpha_v} |G_n \rangle \\
	&=& \widetilde{\U}_O \bigotimes_{v \in O^C} \langle +_{\alpha_v} |G_n \rangle ,
\label{eq:Uo_extract}
\end{eqnarray}
where
\begin{eqnarray}
\nonumber	\widetilde{\U}_O  &=& \prod_{\{ u,v \} \in E,~ u,v\in O} \tCP_{u;v} \\ 
		&& \times \tCN_{r_1; g_V(v_1) \oplus r_1} ... \tCN_{r_n; g_V(v_n) \oplus r_n}
\end{eqnarray}
acts only on the output vertices.
By performing the SPT for flow in the last step from $\Vp$ to $R_V$ on $G_n$, the map given by Eq.(\ref{eq:Uo_extract}) becomes
\begin{eqnarray}
\nonumber	\U_O \U_{spt} \bigotimes_{v \in O^C \backslash \Vp} \langle +_{\alpha_v} |G_n' \rangle,
\end{eqnarray}
where $\U_{spt}$ is a unitary transformation from the space of vertices in $\Vp$ to those in $R_V$, explicitly given by
\begin{eqnarray}
\nonumber	\U_{spt} &=& \J(\alpha_{v_n}) \CZ_{v_n;Odd_{G'} \left( g_V (v_n) \right)} \J(\alpha_{v_{n-1}})... \\
\nonumber		&& ...  \CZ_{v_2;Odd_{G'} \left( g_V (v_2) \right)} \J(\alpha_{v_1}),
\end{eqnarray}
and $G_n'$ is a subgraph of $G_n$ on which $R_V$ is reduced.
Note that $\U_O \U_{spt}$ is a unitary transformation acting on the qubits represented by the wires labeled by vertices in $\Vp$.

The circuit decomposition of the unitary transformation implemented by the last step $\Vp \rightarrow R_V$ is just $\U_O \U_{spt}$.
\begin{figure}
	\centering
	\includegraphics[width=7cm]{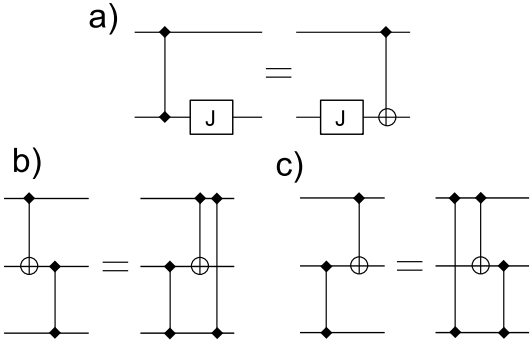}
	\caption{Circuit identities used for the translation.}
	\label{fig:circuit_identity}
\end{figure}
\begin{figure}
	\centering
	\includegraphics[width=6cm]{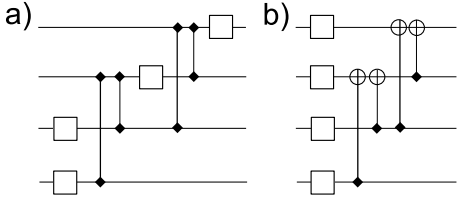}
	\caption{(a) A circuit decomposition obtained by applying the SPT on the graph represented by FIG.~\ref{fig:graphGn}.
		 (b) An equivalent circuit decomposition obtained after applying the circuit identity shown in FIG.~\ref{fig:circuit_identity}(a).}
	\label{fig:Uspt}
\end{figure}
From Lemma~\ref{removeOUT}, the open graph $(G_n', I \backslash R_V, \Vp \cap (O \backslash R_V))$ has a maximally delayed gflow with the partial order $\prec$.
By performing the same manipulations layer by layer, we obtain a total circuit decomposition.

\section{Parallelizing $\J$-gates}\label{sec:parallelizing}
Using the circuit identity presented in FIG.~\ref{fig:circuit_identity}(a), $\U_{spt}$ is transformed into
\begin{eqnarray}
\nonumber	\U_{spt} &=& \bigotimes_{i=0,..,n} \J(\alpha_{v_i}) \CX_{v_n;Odd_{G'} \left( g_V (v_n) \right)} ... \\
\nonumber		&&... \CX_{v_2;Odd_{G'} \left( g_V (v_2) \right)},
\end{eqnarray}
which has the parallelized form for $\J$-gates.
The total unitary transformation $\U$ implemented by the measurement pattern is now written in the form
\begin{eqnarray}
	\U = \U_O \J_0 \U_{\Vp} \J_1 ... \U_{V_d^{\prec}} \J_d,
\label{eq:unitary}
\end{eqnarray}
where each $\U_{V_k^{\prec}}~(k=0,...,d)(V_0^{\prec}=O)$ consists of 2-qubit Clifford gates and each $\J_i$ consists of parallel $\J$-gates.

The circuit representation of $\U$ given by Eq.~(\ref{eq:unitary}) shows that the quantum depth calculated by gflow is lower bounded by the depth calculated by a quantum circuit model that implements all Clifford gates in constant number of steps~\cite{complexity-of-MBQC}.
Each of the $\U_{V_k^{\prec}}~(k=0,...,d)(V_0^{\prec}=O)$ is implemented in constant time by this version of the quantum circuit model.
Each unitary transformation $\J_k~ (k=0,...,d)$ is also implemented in constant time, because all $\J$-gates act on different wires and thus they are parallelized.
Therefore, the total unitary transformation $\U$ represented by Eq.~(\ref{eq:unitary}) is implemented in $c \ast d$ steps by this quantum circuit model, where $c$ denotes a constant.

In \cite{global-circuit-optimization}, a quantum circuit representing a unitary transformation implemented by a measurement pattern on a graph with flow is transformed so that the single-qubit elementary gates are parallelized.
Our method is a generalization of that method for graphs with gflow.

\section{Star pattern transformation for gflow}\label{sec:SPTforGflow}
We define a generalization of the SPT on the graph with gflow but no flow in this section.
A straightforward application of the SPT on a graph with a path cover but without flow does not lead to a well-defined circuit, as we have noted in Sec.\ref{cf}.
The restriction that the target side and the control side of a $\CZ$-gate must act on the same time slice in the circuit prohibits us to write a well-defined circuit for gflow.

We formally define {\it an acausal $\CZ$-gate} as a two-qubit gate acting on two time slices in a circuit.
We denote such an acausal $\CZ$-gate by $\acaus_{u!;v!}$, where the positions $u!$ and $v!$ on which the gate acts may be in different time slices.
We pose two assumptions on the acausal $\CZ$-gate.
First, if the positions $u!$ and $v!$ are regarded to be in the same time slice on the circuit, $\acaus_{u!;v!}$ implements the same map to $\CZ_{u!;v!}$.
Second, if several acausal $\CZ$-gates are acting on the same position $v!$, the map implemented by these gates does not depend on the ordering of the acausal $\CZ$-gates.
The latter assumption originates from the commutativity of the $\CZ$-gates acting on the same physical qubit $v$ on the graph.

\begin{figure}
	\centering
	\includegraphics[width=5cm]{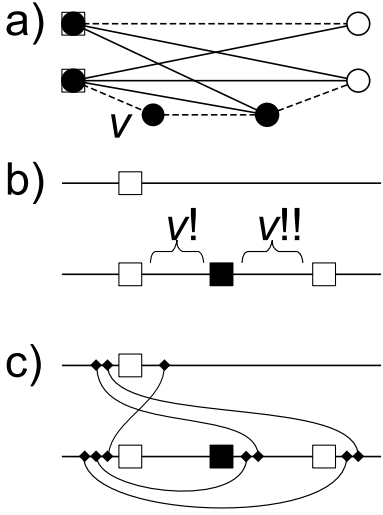}
	\caption{(a) An open graph with gflow but no flow. The set of dashed edges represents the path cover and a vertex is labeled by $v$.
		(b) A quantum circuit given by process (ii) of the SPT (Sec.~\ref{cf}) applied on the graph given by (a). The two wires correspond to the path cover, and boxes represent the $\J$-gates. The gate $\J(\alpha_v)$ is represented by the black box. The position $v!$ is the region of the wire between $\J(\alpha_v)$ and $\J(\alpha_{h^{-1}(v)})$. The position $v!!$ is between $\J(\alpha_v)$ and $\J(\alpha_{h(v)})$.
		(c) A quantum circuit representing a measurement pattern on the graph given by (a).}
	\label{fig:SPT_gflow}
\end{figure}

We refer to a quantum circuit decomposition composed of $\J$-gates, $\CZ$-gates, and acausal $\CZ$-gates as an {\it acausal circuit decomposition}~\cite{gflow}  (see FIG.~\ref{fig:SPT_gflow}(c)).	The SPT on the graph with gflow but no flow is defined to be a procedure to write an acausal circuit representation for a measurement pattern on the graph.
The procedure starts from applying the same processes ($i$) and ($ii$) presented for the SPT for flow in Sec.~\ref{cf}.
These processes are applicable because there is a path cover on any graph with gflow.
Positions are also defined similarly.
The position between $\J(\alpha_{h^{-1}(v)})$ and $\J(\alpha_{v})$ in the wire $v$ is labeled by $v!$, where $h$ is the successor function.
If $v$ is a starting vertex of the path cover, the label $v!$ represents the position before the gate $\J(\alpha_v)$ in wire $v$.
We denote the position between $\J(\alpha_v)$ and $\J(\alpha_{h(v)})$ in wire $v$ by $v!!$, for later use (see FIG.~\ref{fig:SPT_gflow}(b)).
The procedure ends with placing acausal $\CZ$-gates in the circuit instead of the ordinary $\CZ$-gates in the process ($iii$) presented in Sec.~\ref{cf}.
It is always possible to place $\acaus_{u!;v!}$ for any pair of positions $u!$ and $v!$.

\section{Transforming acausal circuits}\label{sec:acausal_circuit}

Although direct application of the SPT on a graph without flow leads to an acausal circuit, our translation method presented in Sec.~\ref{sec:translation} and the SPT have two common aspects.
One is the correspondence between a path cover and a wire of the circuit.  Another is the correspondence between each measurement and a $\J$-gate.
Since our method translates a graph with gflow into a well defined circuit, we expect that the acausal circuit obtained by the SPT should be transformable into a well defined one by taking a suitable circuit transformation.
In this section, we present this circuit transformation.

We define an acausal $\CZ$-gate using ancilla qubits and post-selection to be consistent with the unitary transformation implemented by a measurement pattern with gflow.
In \cite{simulating-CTC-by-MBQC}, an acausal gate is identified with a circuit simulating the effect of closed timelike curve (CTC) (see FIG.~\ref{fig:acausal_CZ_BSS}).
This circuit including ancilla qubits and post-selection of the measurement results is proposed by Bennett and Schumacher \cite{CTCBS} and by Svetlichny \cite{CTCS} to simulate the disordered time effect of CTC by quantum circuits and is called the BSS-type CTC.   In a similar manner, we define acausal gates by using ancilla qubits and post-selection.

\begin{figure}
	\centering
	\includegraphics[width=6cm]{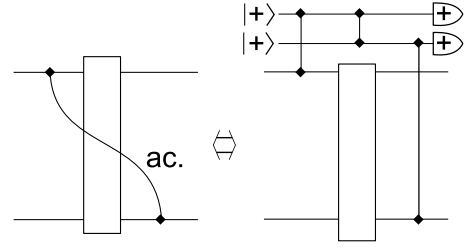}
	\caption{The definition of an acausal $\CZ$-gate. Two ancilla qubits are initially prepared in $| + \rangle$ states, and are post-selected to be in the $| + \rangle$ state at the final measurements.}
	\label{fig:acausal_CZ_def}
\end{figure}

\begin{figure}
	\centering
	\includegraphics[width=6cm]{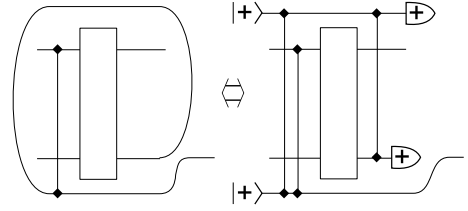}
	\caption{The equivalent circuit representations for the acausal $\CZ$-gates given by FIG.~\ref{fig:acausal_CZ_def}. The proof of the equivalence of the circuits presented in the right hand side and in the right hand side circuit of FIG.~\ref{fig:acausal_CZ_def} is given in Appendix \ref{sec:append}.}
	\label{fig:acausal_CZ_BSS}
\end{figure}

\begin{lem}\label{lem:acausal_valid}
	Consider an acausal circuit obtained by directly applying the SPT on a graph with gflow.
	Define an acausal $\CZ$-gate $\acaus_{u!;v!}$ acting on positions $u!$ and $v!$ by
	\begin{eqnarray}
		\nonumber	&& \acaus_{u!;v!} := \\
			&& \langle +|_{u'} \langle +|_{v'} \CZ_{u!;u'} \CZ_{u';v'} \CZ_{v';v!} |+ \rangle_{u'} |+ \rangle_{v'},
			\label{eq:defacausal}
	\end{eqnarray}
	where $\left \vert {+} \right \rangle_{u'}$ and $\left \vert {+} \right \rangle_{v'}$ represent the initial states of the ancilla qubits, and $\left \langle {+} \right \vert_{u'}$ and $\left \langle {+} \right \vert_{v'}$ represent a post selected measurement branch for a projective measurement described by $\{ | \pm \rangle_{i'} \langle \pm | \}$ for $i=u,v$, respectively. We post-select the measurement result ``$+$''. (See FIG.~\ref{fig:acausal_CZ_def}.)
	Then the acausal circuit represents a unitary transformation equivalent to the one implemented by the measurement pattern on the graph.
\end{lem}

$Proof$)
Rewriting the acausal $\CZ$-gates according to Eq.~(\ref{eq:defacausal}) is always possible.
This is because all the $\CZ$-gates appearing in Eq.~(\ref{eq:defacausal}) commute with each other, and there are no other gates which define an ordering of gates on the ancilla qubits $u'$ and $v'$.

Using the definition of an acausal $\CZ$-gate given by Eq.~(\ref{eq:defacausal}), the acausal circuit can be transformed into a well-defined one.
The circuit is composed of three parts, preparation of initial ancilla states, a circuit consisting of $\J$-gates and $\CZ$-gates, and final measurements.
The circuit in the second part can be transformed to a measurement pattern by performing the inverse transformation of the SPT \cite{flow-to-circuit}.  Thus we obtain the corresponding open graph $(G',I',O')$ with flow given by

\begin{eqnarray}
		\nonumber V(G') &=& V(G) \cup_{\{ u,v \} \notin P_h} \{ u' ,v' \},\\
		\nonumber E(G') &=& E(G) \cup_{\{ u,v \} \notin P_h} \{ E_{u u'} E_{u' v'} E_{v' v} \} \\
		\nonumber	 &&  \backslash \cup_{\{ u,v \} \notin P_h} E_{u v},\\
		\nonumber I' &=& I \cup_{\{ u,v \} \notin P_h} \{ u' , v' \}, \\
		\nonumber O' &=& O \cup_{\{ u,v \} \notin P_h} \{ u' , v' \},
\end{eqnarray}
where $P_h$ is defined by substituting the successor function $h$ instead of the flow function used in Eq.~(\ref{eq:pathcover}).
There are trivial I-O paths from the newly added vertices to themselves since they are included both in the input set and in the the output.
The original path cover on the open graph $G$ and these trivial paths construct a path cover on the open graph $G'$.
Let us denote the unitary transformation implemented by $(G',I',O')$ by $\U_{G'}$.
Then from $\U_{G'} \propto \bigotimes_{v \in O^C} \langle +_{\alpha_v} |G' \rangle$, we have
\begin{eqnarray}
	\nonumber	\bigotimes_{\{ u,v \} \notin P_h} \langle +|_{u'} \langle +|_{v'} \U_{G'} |+ \rangle_{u'} |+ \rangle_{v'} \\
	\propto \bigotimes_{\{ u,v \} \notin P_h} \langle +|_{u'} \langle +|_{v'} \bigotimes_{v \in O^C} \langle +_{\alpha_v} |G' \rangle |+ \rangle_{u'} |+ \rangle_{v'}.
	\label{eq:yoiyoi}
\end{eqnarray}
Since all the projectors commute, we can perform the projector $|+ \rangle \langle +|$ on the ancilla qubits first.
Using the relation
\begin{eqnarray}
\nonumber	\tCP_{u;v} \propto \langle +|_{u'} \langle +|_{v'} \tCP_{u;u'} \tCP_{u';v'} \tCP_{v';v} |+ \rangle_{u'} |+ \rangle_{v'} \\
\nonumber	(\forall \{ u,v \} \notin P_h),
\label{eq:decompCZ}
\end{eqnarray}
we have
\begin{eqnarray}
	\nonumber	&& \bigotimes_{\{ u,v \} \notin P_h} \langle +|_{u'} \langle +|_{v'} \bigotimes_{v \in O^C} \langle +_{\alpha_v} |G' \rangle |+ \rangle_{u'} |+ \rangle_{v'} \\
	\nonumber		&\propto& \bigotimes_{\{ u,v \} \notin P_h} \langle +|_{u'} \langle +|_{v'} \\
	\nonumber	&& \bigotimes_{v \in O^C} \langle +_{\alpha_v}| \tCP_{u;u'} \tCP_{u';v'} \tCP_{v';v} |G' \cap P_h \rangle |+ \rangle_{u'} |+ \rangle_{v'} \\
	\nonumber	&\propto& \bigotimes_{v \in O^C} \langle +_{\alpha_v} |G \rangle ,
\end{eqnarray}
namely, the graph $G'$ now returns to the original graph $G$.
Thus if the ancilla qubits of $\U_{G'}$ are prepared in $|+ \rangle$ and the final measurements post-select the final states to be $| + \rangle$, the unitary transformation implemented by the measurement pattern on $(G,I,O)$ is also implemented in this post-selected way.\qed

The definition of an acausal gate by Eq.~(\ref{eq:defacausal}) is equivalent to the BSS-type CTC \cite{CTCBS, CTCS} (FIG.~\ref{fig:acausal_CZ_BSS}).
The $\CZ$-gate appearing on the left side circuit of FIG.~\ref{fig:acausal_CZ_BSS} acts on two-qubits where one of the qubits has returned from a future to its past.
Acausal $\CZ$-gates defined by the circuit presented on the left side picture of FIG.~\ref{fig:acausal_CZ_BSS} also appear in Ref.~\cite{simulating-CTC-by-MBQC}.

Although the acausal circuit is transformed into an ordinary circuit without acausal gates, it cannot be implemented deterministically since the circuit includes post-selection of measurement results.
However we will show that it is possible to transform the circuit to an ordinary deterministically implementable circuit by taking further transformations.

Next lemma shows that a further transformation on the acausal circuit equivalent to the transformation we have applied on the graph in Eq.~(\ref{eq:change}) is possible.
\begin{lem}\label{lem:extend_change}
	Let us denote an acausal circuit obtained by applying the SPT on an open graph $G$ by $C_{spt}^G$.
	Then the following identity holds for a sequence of open graphs $\{ G_i \}$ defined by Eq.~(\ref{eq:def_sequence_G}) up to a normalization factor:
	\begin{eqnarray}
		\CX_{h(v_i);g_V(v_i) \oplus h(v_i)} C_{spt}^{G_{i-1}} = C_{spt}^{G_i}.
	\label{eq:acausal_change}
	\end{eqnarray}
\end{lem}

\begin{figure}
	\centering
	\includegraphics[width=8cm]{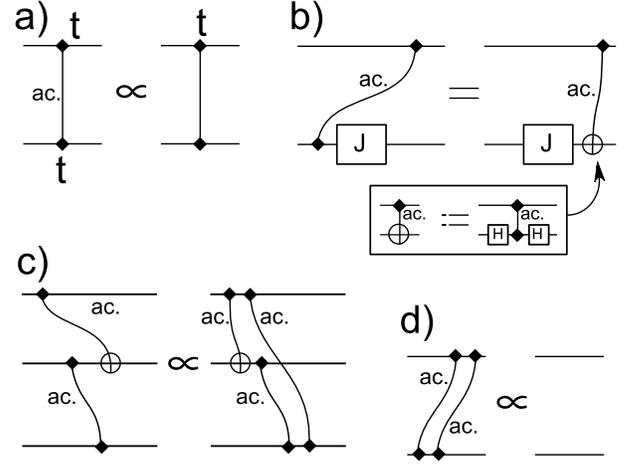}
	\caption{The circuit identity satisfied by the acausal gates. The acausal gates are labeled ``ac.'' The acausal CNOT-gate is defined as the acausal $\CZ$-gate sandwiched by Hadamard gates. The proof of these identities is given in Appendix \ref{sec:append}}.
	\label{fig:acausal_CZ_identity}
\end{figure}

$Proof$)
We first replace the CNOT-gates appearing on the left hand side of Eq.~(\ref{eq:acausal_change}) with acausal CNOT-gates.
We shift the position of the target of these acausal CNOT-gates backwards in time, until all the acausal CNOT-gates are changed to acausal $\CZ$-gates by the circuit identity presented in FIG.~\ref{fig:acausal_CZ_identity}(b).

The target of a CNOT-gate acting on $v(\in g_V(v_i) \oplus h(v_i))$ first hits the (acausal) $\CZ$-gates corresponding to non-path edges incident from the vertex $v$.
\begin{description}
	\item[process 1] By commuting the acausal CNOT-gate and these (acausal) $\CZ$-gates, new (acausal) $\CZ$-gates are created between $h(v_i)$ and $N(v) \backslash h^{-1} (v)$, using the circuit identity presented in FIG.~\ref{fig:acausal_CZ_identity}(c).
	\item[process 2] Further commuting the acausal CNOT-gate with $\J(\alpha_{v})$, it changes to an acausal $\CZ$-gate using the circuit identity presented in FIG.~\ref{fig:acausal_CZ_identity}(b).
\end{description}
By {\bf process 1} and {\bf process 2}, the acausal CNOT-gate is transformed to
\begin{eqnarray}
\nonumber	\acaus_{h(v_i)!!;N(v) \backslash h^{-1} (v)!!} \acaus_{h(v_i)!!;h^{-1} (v)!!} \\
\nonumber	\propto \acaus_{h(v_i)!!;N(v)!!}.
\end{eqnarray}
This holds since two acausal $\CZ$-gates acting on the same pair of positions are canceled by the circuit identity presented in FIG.~\ref{fig:acausal_CZ_identity}(d).
If we perform these transformations for all the CNOT-gates appearing on the left hand side of Eq.~(\ref{eq:acausal_change}), the CNOT-gates transform
\begin{eqnarray}
\nonumber	\prod_{v \in g_V(v_i) \oplus h(v_i)} \acaus_{h(v_i)!!;N(v)!!} \\
\nonumber	= \prod_{v \in g_V(v_i)} \acaus_{h(v_i)!!;N(v)!!} \acaus_{h(v_i)!!;N(h(v_i))!!} \\
\nonumber	\propto \acaus_{h(v_i)!!;Odd (g_V(v_i)) \backslash v_i!!} \acaus_{h(v_i)!!;N(h(v_i)) \backslash v_i!!}.\\
	\label{eq:acausal_graph}
\end{eqnarray}
The final line holds since even number of acausal $\CZ$-gates acting on the same pairs of positions are canceled by the circuit identity presented in FIG.~\ref{fig:acausal_CZ_identity}(d).
The last term  $\acaus_{h(v_i)!!;N(h(v_i)) \backslash v_i!!}$ cancels all (acausal) $\CZ$-gates incident from $v_i !!$.
New (acausal) $\CZ$-gates are created by the first term $\acaus_{h(v_i)!!;Odd (g_V(v_i)) \backslash v_i!!}$.
These transformations directly correspond to the transformation from $G_{i-1}$ to $G_i$ given by Eq.~(\ref{eq:def_sequence_G}).
\qed

\begin{figure}
	\centering
	\includegraphics[width=6cm]{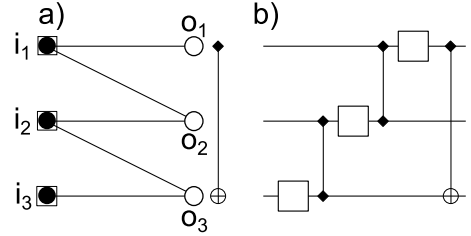}
	\caption{(a) An open graph transformed according to Eq.~(\ref{eq:change}) from the graph given by FIG.~\ref{fig:gYfN} . The extra CNOT-gate must act on the qubits $o_1$ and $o_2$.
		 (b) The circuit decomposition obtained by our method in Sec.~\ref{sec:translation}.}
	\label{fig:gYfN_trans}
\end{figure}

\begin{figure}
	\centering
	\includegraphics[width=7cm]{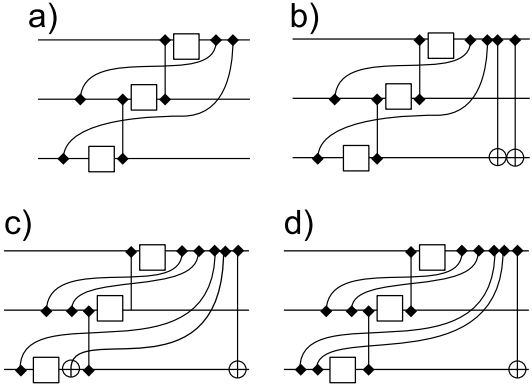}
	\caption{(a) The acausal circuit representation for a unitary transformation implemented by the measurement pattern on the open graph presented in FIG.~\ref{fig:gYfN}. We choose the edges $(i_2,o_1)$ and $(i_3,o_1)$ as the acausal $\CZ$-gates.
		 (b) Applying $I = CNOT \cdot CNOT$ on the control qubit $h(i_1) = o_1$, and the target qubit $g_V(i_1) \backslash h(i_1) = \{ o_3 \}$.
		 (c) Shifting the position of the acausal CNOT-gate before the $\J$-gate through the $\CZ$-gate.
		 (d) Shifting the position of the acausal CNOT-gate through the $\J$-gate. It is changed to an acausal $\CZ$-gate.
		 By cancelling the pairs of acausal $\CZ$-gates according to the circuit identity given in FIG.~\ref{fig:acausal_CZ_identity}(d), this acausal circuit is transformed to the circuit presented in FIG.~\ref{fig:gYfN_trans}(b).}
	\label{fig:gYfN_acausal}
\end{figure}

Now we present how to transform acausal circuits into ordinary circuits by using Lemma~\ref{lem:extend_change}.	An example is shown in FIG.~\ref{fig:gYfN_acausal} where an acausal circuit obtained from the open graph presented in FIG.~\ref{fig:gYfN} is transformed to an ordinary circuit presented in FIG.~\ref{fig:gYfN_trans}(b).
Since a sequence of two CNOT-gates acting on the same qubits are equivalent to an identity gate,
\begin{eqnarray}
\nonumber	&& C_{spt}^{G_{i-1}} \\
\nonumber	&=& \CX_{h(v_i);g_V(v_i) \oplus h(v_i)} \CX_{h(v_i);g_V(v_i) \oplus h(v_i)} C_{spt}^{G_{i-1}} \\
\nonumber	&=& \CX_{h(v_i);g_V(v_i) \oplus h(v_i)} C_{spt}^{G_i}. \\
\end{eqnarray}
We start from $C_{spt}^{G'}$ and repeat this transformation.
We finally obtain an acausal circuit for $C_{spt}^{G_n}$ followed by a set of CNOT-gates:
\begin{eqnarray}
\nonumber	C_{spt}^{G'} &=& \CX_{h(v_1);g_V(v_1) \oplus h(v_1)} C_{spt}^{G_1} \\
\nonumber		&=& ... \\
\nonumber		&=& \CX_{g_V(v_1) \oplus h(v_1)} ... \CX_{h(v_n);g_V(v_n) \oplus h(v_n)} C_{spt}^{G_n}.\\
\end{eqnarray}
Since there exists flow between the last two layers of $G_n$, all acausal $\CZ$-gates in $C_{spt}^{G_n}$ can be changed into ordinary $\CZ$-gates (FIG.~\ref{fig:acausal_CZ_identity}(a)).
The unitary map from $\Vp$ to $O$ is now written as $\U_O \U_{spt}$.
There is no acausal $\CZ$-gate connecting a position in $\U_{spt}$ and a position in the acausal circuit of the following layer, therefore we can perform the same transformation on the following layer independently.

Lemma~\ref{lem:extend_change} shows the equivalence between the transformations on a graph and that on the acausal circuit.
If two $\CZ$-gates act on the same pair of vertices on the graph, they cancel each other.
The analogue of this cancelation on the acausal circuit is the identity presented in FIG.~\ref{fig:acausal_CZ_identity}(d).
Despite the fact that we have defined the acausal circuit including ancillas and post-selection, the transformation shows that all acausal gates cancel in this case and deterministic implementation of a unitary transformation is possible.

\section{Depth compression and acausal circuits}\label{sec:depth}
In Section \ref{sec:parallelizing}, we have shown that the unitary transformation implemented by the measurements on qubits in a single layer of gflow is written as a parallelized $\J$-gates followed by a sequence of Clifford gates.
Any unitary transformation that is written in this form is implemented in a constant quantum depth by the measurement pattern. This is in contrast to a variation of quantum circuit model where classically controlled operations depending on measurement outcomes are not included. Generally, in such a model, the quantum depth depends on the system size.

In this section, we show how the acausal circuit obtained by directly applying the SPT expresses the depth compression by extending the definition of the temporal ordering of gates.	The acausal gates are regarded as shorthands for circuits implementing gate-teleportation \cite{gate-teleportation}, in this section.
With the aid of ancilla qubits and post-selection, this circuit can be interpreted to have a power equivalent to sending quantum states back into the past, from where the computation continues again.
The condition of post-selection is circumvented by applying suitable correction operators depending on the outcomes of the Bell measurement in gate-teleportation.

\subsection{Gate-teleportation}
\begin{figure}
	\centering
	\includegraphics[width=6cm]{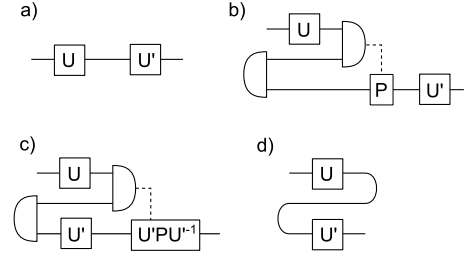}
	\caption{(a) A circuit to implement a unitary transformation $\U'\U$. The unitary transformations $\U$ and $\U$ may consist of sequences of gates. (b) A circuit obtained by inserting a teleportation between $\U$ and $\U'$ presented in (a). The Half-ellipsoids at the left and right end of wires represent a preparation of maximally entangled state and a Bell measurement, respectively.	The white box labeled ``P'' represents the Pauli correction operator.	This is a circuit representation of the sequence of operations presented in Eq.~(\ref{eq:teleportation}).	(c) The circuit representation of the sequence of operations presented in Eq.~(\ref{eq:postponing correction}). (d) A circuit obtained by assuming post-selection in the Bell measurement in the circuit presented in (c). The curved wire is an abbreviation of the post-selected teleportation similarly to the curved wire presented in Fig.~\ref{fig:acausal_CZ_BSS}.	This abbreviation is motivated by the fact that quantum states can be sent ``back in time'' by post-selected teleportation. }
	\label{fig:gate-teleportation}
\end{figure}

Let us describe how to parallelize unitary transformations and probabilistically compress the quantum depth by using a post-selection version of the gate-teleportation protocol \cite{gate-teleportation}.	We first review the gate-teleportation protocol for this purpose.	Consider a one-qubit circuit representing two unitary transformations $\U$ and $\U'$ applied on a single Hilbert space labeled $A$ in a sequence (see Fig.~\ref{fig:gate-teleportation}(a)).	The unitary transformation $\U'_A$ must be applied after the gate $\U_A$ for implementing a unitary transformation $\U' \U$ on a state on $A$.

With the aid of spatial resources, namely, ancilla qubits, we construct another circuit where $\U$ and $\U'$ are applied on different  qubits {\it in parallel} but the {\it ordered sequence} of unitary transformations $\U' \U$ is still implemented.	Consider a circuit that has three-qubits labeled by $A$, $B$ and $C$, depicted in Fig.~\ref{fig:gate-teleportation}(b). The initial state of qubits $B$ and $C$ is prepared in a maximally entangled state $\ket{\Psi}_{BC} := \CZ_{BC} |+\rangle_B |+\rangle_C$. After performing unitary transformation $\U$ on qubit $A$, qubits $A$ and $B$ are measured in a basis $\{ \CZ | s_A \rangle_A |s_B \rangle_B \}_{s_A, s_B \in \{+,-\}}$. This measurement is called the Bell measurement. A Pauli correction operator $\Pau_C(s_A,s_B)$ depending on the measurement outcome is applied on qubit $C$, where $\Pau_C(+,+)=\I_C$ (do nothing), $\Pau_C(+,-) = \X_C$, $\Pau_C(-,-) = \Y_C$ and $\Pau_C(-,+)=\Z_C$.	Finally the unitary transformation $\U'$ is performed on qubit $C$. In short, after $\U_A$, we perform a quantum teleportation from $A$ to $C$, followed by $\U'_C$.	An initial state $\ket{\phi}_A \ket{\Psi}_{BC}$ is transformed to
\begin{eqnarray}
	\U'_C \Pau_C(s_A,s_B) \bra{s_A} \bra{s_B} \U_A \ket{\Psi}_{BC} \ket{\phi}_A = \frac{1}{2} \U'_C \U_C \ket{\phi}_C,\nonumber \\
	\label{eq:teleportation}
\end{eqnarray}
namely, apart from the difference on the Hilbert space where the output state is obtained, the circuit presented in Fig.~\ref{fig:gate-teleportation}(b) implements the same unitary transformation as the circuit presented in Fig.~\ref{fig:gate-teleportation}(a).

In this form, we still have to perform $\U'_C$ after $\U_A$, since $P'_C(s_A,s_B)$ must be performed after the outcome $(s_A,s_B)$ is obtained by the measurement on qubits $A$ and $B$.	However, we can rewrite (\ref{eq:teleportation}) as
\begin{eqnarray}
	&&\textrm{L. H. S of }(\ref{eq:teleportation}) = \nonumber\\
	&& \U'_C \Pau_C(s_A,s_B) \U'^{-1}_C   \bra{s_A} \bra{s_B} \CZ_{AB} \U'_C \U_A \ket{\Psi}_{BC} \ket{\phi}_A. \nonumber \\
	\label{eq:postponing correction}
\end{eqnarray}
On the right hand side of Eq.~(\ref{eq:postponing correction}), $\U_A$ and $\U'_C$ is applied in parallel, before the correction operator $\U'_C \Pau_C(s_A,s_B) \U'^{-1}_C$ (see Fig.~\ref{fig:gate-teleportation}(c)).	In general, the quantum depth for implementing $\U'_C \Pau_C(s_A,s_B) \U'^{-1}_C$ may be greater than that for implementing $\U'_C$.	Thus the quantum depth is not necessarily reduced in spite of the unitary transformations $\U$ and $\U'$ are parallelized.

If the measurement result is post-selected to give $(s_A,s_B) = (+,+)$, there is no need for the correction, and the quantum depth for implementing $\U'\U$ is reduced by parallelizing $\U$ and $\U'$.	In this post-selected branch, the teleportation represents a map that can be interpreted to send a quantum state  ``back in time", in the same way as in the interpretation of the BSS-type CTC \cite{CTCBS, CTCS}.  The argument of this interpretation is given as follows.   Consider a case where $\U = \U' =\I$ and we perform a measurement in an arbitrary basis $\{ \bra{\varphi}_m \}_{m \in M}$ on the output state obtained in $C$ {\it before the input state $\ket{\phi}_A$ is prepared}. The probability to obtain measurement result $m'$ is
\begin{eqnarray}
	&& 4 || \bra{+}_A \bra{+}_B \CZ_{AB} \ket{\phi}_A \bra{\varphi}_{m'} \ket{\Psi}_{BC}||^2 \nonumber\\
	&&= 4 || \bra{\varphi}_{m'} \bra{+}_A \bra{+}_B \CZ_{AB} \ket{\phi}_A \ket{\Psi}_{BC}||^2 \nonumber\\
	&&= ||\bra{\varphi}_{m'} \ket{\phi}_A  ||,^2
	\label{eq:sent-back}
\end{eqnarray}
in the post-selected branch. This equality implies that the probability distribution of the measurement performed on $C$ before the preparation of state $\ket{\phi}_A$ is equal to the probability distribution of the measurement performed on state $\ket{\phi}_A$.	This is equivalent to saying that the state $\ket{\phi}_A$ is sent back in time, since the basis of the measurement is arbitrary. From this perspective, the teleportation protocol can be interpreted to reduce the quantum depth by sending the quantum state into the past and allowing the computation to continue again from there.

Using the notation introduced for representing the BSS-type CTC, a circuit for sending a quantum state back in time should be depicted by a curved wire as represented in Fig.~\ref{fig:gate-teleportation}(d).	Note that this expression indicates that the Hilbert spaces described by qubits $A$ and $C$ are considered to be the Hilbert spaces of an identical qubit at two different {\it temporal} positions, instead of the Hilbert spaces of the two {\it spatially} different qubits existing at the same time.

\begin{figure}
	\centering
	\includegraphics[width=6cm]{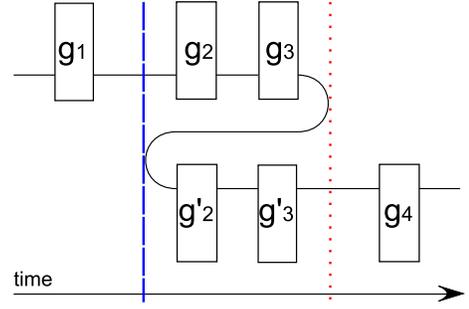}
	\caption{(color online) A circuit which has extended temporal ordering of gates. The white boxes labeled by $g_i$ represent quantum gates. At the time represented by a dotted line, the state is sent back in to the time represented by a dashed line. The partial ordering of gates in this circuit is defined by $g_1 \leq g_2 \leq g_3 \leq g_4$ and $g_1 \leq g'_2 \leq g'_3 \leq g_4$.}
	\label{fig:partial_order_curve}
\end{figure}

The definition of the partial ordering of gates included in a circuit with such curved wires must be extended from that defined on an ordinary circuit. In a circuit without curved wires, the partial ordering $g_1 \leq g_2$ between gates $g_1$ and $g_2$ applied on the same qubit is defined when $g_2$ is performed {\it after} $g_1$. Quantum depth of a circuit without curved wires is then defined as the the maximum number of gates included in any sequence of gates $g_1,~g_2,~...,~g_n$ such that $g_i \leq g_{i+1}$.   To define the partial ordering on gates included in circuits with curved wires, we have to reconsider how to define a situation to describe a gate $g_2$ is performed {\it after} $g_1$.   Consider a circuit depicted in Fig.~\ref{fig:partial_order_curve}. Three gates $g_1$, $g_2$ and $g_3$ are performed sequentially, and the state is sent back from the time represented by a dotted line to the time represented by a dashed line, and then gates $g'_2$, $g'_3$ and $g_4$ are applied in a sequence. Partial orderings $g_1 \leq g_2 \leq g_3$ and $g'_2 \leq g'_3 \leq g_4$ should be defined as usual. Let us assume the time represented by the dotted line is just after $g_3$ and just before $g_4$, and the time represented by the dashed line is just after $g_1$ and just before $g'_2$. Thus partial orderings $g_1 \leq g'_2$ and $g_3 \leq g_4$ are defined by using the time represented by the dashed and the dotted lines as intermediaries, respectively.  However there should be no partial ordering between $g_3$ and $g'_2$. Extending the definition of partial ordering in this way, quantum depth of a circuit with curved wires is defined in the same way to that for a circuit without curved wires.
  
Parallelization of unitary transformations is to remove the partial ordering between the unitary transformations by using curved wires. From this perspective, the teleportation protocol reduces the quantum depth by extending the definition of temporal ordering of gates.

Of course post-selection is probabilistic. The teleportation protocol for sending the quantum state back in time, or equivalently, the operation represented by the circuit with curved wires, cannot be deterministically implemented in quantum mechanics in general. However in some cases, we can deterministically {\it simulate} it by applying appropriate correction operators.	If the correction $\U'_C \Pau'_C(s) \U'^{-1}_C$ is implementable more efficiently than applying $\U'$ or $\U$, it is possible to reduce the quantum depth without assuming post-selection.	For example, if $\U'$ consists of a sequence of Clifford gates, $\U'_C \Pau'_C(s) \U'^{-1}_C$ is equivalent to a Pauli operator whose quantum depth is one.	Thus the quantum depth for implementing the circuit obtained by parallelizing $\J$-gates as in Sec.~\ref{sec:parallelizing} can be reduced to $c*d$, where $c$ and $d$ denotes a constant and the depth of gflow, respectively.	As we will see in the next subsection, there are non-Clifford unitary transformations that change certain Pauli correction operators into other Pauli operators, and these transformations are relevant to the quantum depth of patterns with gflow.	Note that the gate-teleportation protocol reduces the quantum depth by using additional ancillary systems and classically controlled operations.

\subsection{Depth compression described by acausal gates}
In this subsection, we apply the mechanism of depth compression presented in the previous subsection to the acausal circuits obtained by directly applying the SPT on measurement patterns with gflow.	It is possible to reduce the quantum depth of certain types of gate sequences including $\J$-gates, which are not Clifford in general.	Pauli $Z$ operator changes to Pauli $X$ operator by the conjugate action of a $\J$-gate.	This allows us to reduce the quantum depth of a circuit decomposition obtained by applying the SPT on the measurement pattern implemented in the last two layers of $G_n$ defined in Eq.~(\ref{eq:def_sequence_G}) (ex. the circuit presented in Fig.~\ref{fig:Uspt}(a)).	Note that the open graph constituting of the vertices of last two layers of $G_n$ has a gflow with its depth one.
\begin{figure}
	\centering
	\includegraphics[width=6cm]{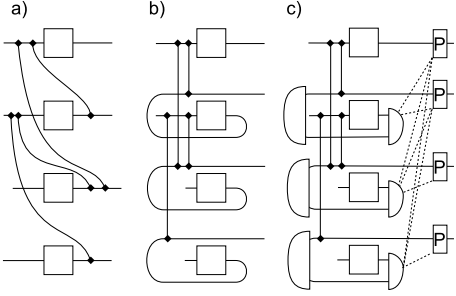}
	\caption{(a) An acausal circuit description of the circuit decomposition presented in Fig.~\ref{fig:Uspt}(a). (b) An equivalent circuit with curved wires obtained after representing the acausal $\CZ$-gates by using a circuit to represent the BSS-type CTC. (c) A circuit to implement the unitary transformation described by the circuit presented in (a) and (b) deterministically, by employing Pauli corrections depending on the measurement outcomes of the Bell measurements. The white boxes labeled by $P$ represent Pauli operators. (The sequence of $\CZ$-gates should be parallelized by other gate-teleportations not depicted here.)}
	\label{fig:zigzag}
\end{figure}

The acausal $\CZ$-gates describe the depth compression of circuits of this type through its equivalent description by the curved wires.	Consider again the circuit depicted in Fig.~\ref{fig:Uspt}(a).	The acausal circuit description of this circuit presented in FIG.~\ref{fig:zigzag}(a) is equivalent to the circuit with curved wires presented in FIG.~\ref{fig:zigzag}(b), which can be deterministically simulated by performing suitable Pauli corrections as presented in FIG.~\ref{fig:zigzag}(c).   The circuits presented in Fig.~\ref{fig:zigzag}(a), (b) and (c) show that $\J$-gates are applied in parallel by considering the extended temporal ordering of the gates.	Thus circuits with acausal $\CZ$-gates, not only represent the unitary transformation implemented by the original pattern, but also capture the extended temporal ordering of gates after parallelizing gates by the gate-teleportation protocol.

\begin{figure}
	\centering
	\includegraphics[width=6cm]{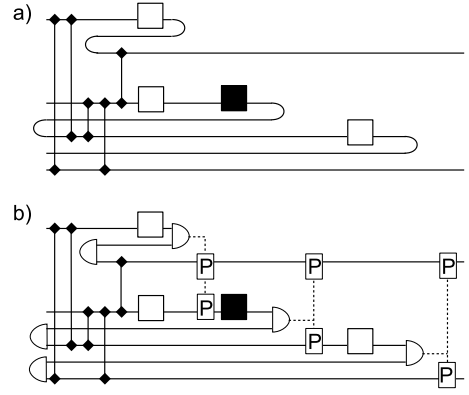}
	\caption{(a) An acausal circuit equivalent to the one presented in Fig.~\ref{fig:SPT_gflow}(c) obtained after representing the acausal $\CZ$-gates by using a circuit to represent the BSS-type CTC. (b) A circuit to implement the unitary transformation described by the circuit presented in (a) deterministically, by employing Pauli corrections depending on the outcomes of the Bell measurements. The white boxes labeled ``P'' represent Pauli correction operators.}
	\label{fig:6vers_correction}
\end{figure}

This example shows that a class of unitary transformations implemented by acausal circuits and circuits with curved wires in constant quantum depth can be deterministically implemented by measurement patterns and circuits simulating the curved wires by gate-teleportation protocol, also in constant quantum depth. Of course this does not necessarily hold for all unitary transformations, since it may take a large quantum depth for applying correction operators in gate-teleportation protocols in general.	The examples presented in Fig.~\ref{fig:zigzag} and \ref{fig:6vers_correction} indicate several unitary transformations and their acausal circuit representations are given by simply applying the direct SPT on measurement patterns with gflow.

\section{Concluding remarks}
In this work, we have constructed two methods to translate a unitary transformation implemented by a measurement pattern on a graph with gflow to its circuit representation to analyze the trade-off relationship of the spatial and temporal resources in MBQC.
The first method is a generalization of the method presented in Ref.~\cite{global-circuit-optimization} and is also an extended version of the method proposed in Ref.~\cite{compact}.
We have divided an open graph with gflow into layers, and transformed each layer into an open graph with flow followed by a sequence of CNOT-gates.
The unitary transformation implemented by the measurements on each layer is thus written by the part obtained by the SPT and the sequence of CNOT-gates.
We also transformed the SPT part into a circuit consisting of a sequence of Clifford gates and parallelized $\J$-gates.
The resulting circuit exhibits that the depth of gflow corresponds to the depth calculated by a special version of quantum circuit model where any sequence of Clifford gates is regarded to be implemented in constant depth.

In the second translation method, a measurement pattern is translated via an acausal circuit including acausal gates obtained by directly applying the SPT on a graph with gflow but no flow.
We defined these acausal gates in terms of post-selection.
Based on this definition, all acausal gates can be canceled by taking appropriate transformations of the circuit.	This leads to a well-defined ordinary circuit representation for the measurement pattern, which is equivalent to the one obtained by the first translation method.

Finally we have shown how the acausal circuits obtained by directly applying the SPT on measurement patterns with gflow express the depth compression by introducing an extended definition of the temporal ordering of gates in their equivalent circuit representation with curved wires.	For deterministically simulating the circuit with curved wires  by using a gate-teleportation protocol, appropriate correction operators depending on the outcomes of the Bell measurements are required.

We conjecture that in order to lift the condition of post-selection entirely from the acausal circuits, it is sufficient to perform a single layer of Pauli corrections per layer of $\J$-gates according to the measurement outcomes for implementing the acausal $\CZ$-gates acting astride the $\J$-gates, as depicted in Fig.~\ref{fig:zigzag}(c) and in Fig.~\ref{fig:6vers_correction}.	This is because any part of an acausal circuit corresponding to a single layer of gflow can be rewritten into a circuit constituting of a single layer of $\J$-gates and a sequence of Clifford gates, and it suffices to perform a single layer of Pauli corrections for parallelizing all the Clifford gates included in the sequence. If this conjecture is true, the quantum depth of the gate-teleportation protocol to implement a gate-sequence corresponding to the measurements on qubits in a single layer of gflow becomes constant.	This implies that all acausal circuits obtained by directly applying the SPT on measurement patterns with gflow can be simulated by the gate-teleportation protocol with the depth equal to the depth of gflow.

Our formulation presents a way to understand the trade-off relationship between the temporal and spatial resources in quantum computation in terms of the extended temporal ordering of the acausal circuits.	The temporal resource, or quantum depth, of quantum computation represented by MBQC  is reduced from the ordinary quantum circuit model by extending the definition of the temporal ordering of gates.	The spatial resource, or an ancilla system, is required for simulating the extended temporal ordering by the gate-teleportation protocol.	If our conjecture is true, this mechanism of the trade-off relationship explains the depth compression of MBQC.	We leave the rigorous study of this conjecture for future works.

\begin{acknowledgments}
We thank D. Markham, K. Nakago and E. Pius for helpful discussions. This work was supported by Project for Developing Innovation Systems of the Ministry of Education, Culture, Sports, Science and Technology (MEXT), Japan.  M.~M.~and M.~H.~acknowledge support from JSPS by KAKENHI  (Grant No. 23540463, No. 23240001, No. 26330006 and No.23-01770). M.~H.~also acknowledges support from Singapore's National Research Foundation and Ministry of Education. This material is based on research funded by the Singapore National Research Foundation under NRF Award NRF-NRFF2013-01. J.~M.~is supported by Leading Graduate Course for Frontiers of Mathematical Sciences and Physics.  The authors also gratefully acknowledge the ELC project (Grant-in-Aid for Scientific Research on Innovative Areas MEXT KAKENHI (Grant No. 24106009)) for encouraging the research presented in this paper.
\end{acknowledgments}

\appendix

\section{Proof of the circuit identities in FIG.~\ref{fig:acausal_CZ_identity} and the equivalence of the circuits in FIG.~\ref{fig:acausal_CZ_def} and FIG.~\ref{fig:acausal_CZ_BSS}.}\label{sec:append}
\begin{figure}
	\centering
	\includegraphics[width=6cm]{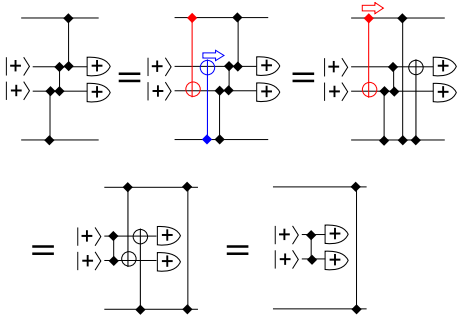}
	\caption{(color online) A proof of the circuit identity presented in  FIG.~\ref{fig:acausal_CZ_identity}(a). The first equality is by Eq.~(\ref{eq:stateCNOT}) and the last equality is by Eq.~(\ref{eq:measCNOT}). The second and the third equalities are by Eq.~(\ref{eq:CNOT_CZ}). The shifts of the position of the CNOT-gates in the second and third equalities are allowed because the acausal gate acts at a single time. The probability to obtain the correct measurement result is always one quarter.}
	\label{fig:pf_acCZ_CZ}
\end{figure}

Once the acausal gates shown in FIG.~\ref{fig:acausal_CZ_identity} are replaced  by ordinary $\CZ$-gates and CNOT-gates, the circuit identities presented in FIG.~\ref{fig:acausal_CZ_identity} are easily seen to hold.
We prove that these identities also hold for acausal gates.
The transformation method of circuits is depicted in FIG.~\ref{fig:pf_acCZ_CZ} to FIG.~\ref{fig:pf_acCNOT_acCZ}.
There are two identities commonly used in the translation method.

First, we can add and remove CNOT-gates when the target qubit is prepared in $|+ \rangle$ or post-selected to the state $|+ \rangle$, namely,
\begin{eqnarray}
	|+ \rangle_B  |rest \rangle_A = \CX_{A;B} |+ \rangle_B  |rest \rangle_A,
\label{eq:stateCNOT}
\end{eqnarray}
\begin{eqnarray}
	\langle +|_B  \langle rest|_A = \langle +|_B  \langle rest|_A \CX_{A;B},
\label{eq:measCNOT}
\end{eqnarray}
where $| rest \rangle $ represents the state outside system $B$ but including system $A$.

Second, we use the following identity relation when shifting positions of CNOT-gates through $\CZ$-gates,
\begin{eqnarray}
	\CZ_{A;B} \CX_{C;B} = \CX_{C;B} \CZ_{A;C} \CZ_{A;B}.
\label{eq:CNOT_CZ}
\end{eqnarray}
This identity relation appears when the acausal CNOT-gate presented in FIG.~\ref{fig:acausal_CZ_identity}(c) is replaced by an ordinary CNOT-gate.

\begin{figure*}
	\centering
	\includegraphics[width=14cm]{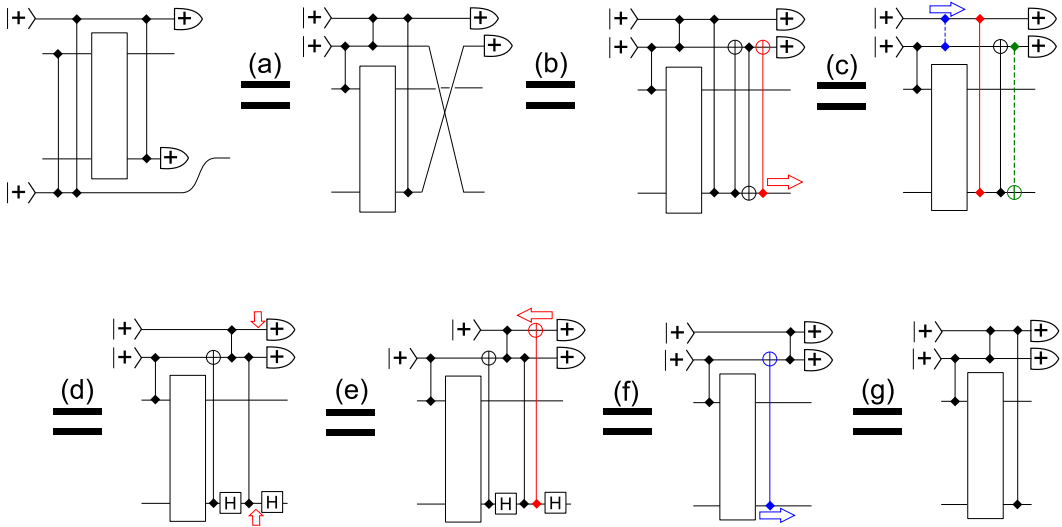}
	\caption{(color online) A proof of the equivalence of the circuits presented in FIG.~\ref{fig:acausal_CZ_def} and FIG.~\ref{fig:acausal_CZ_BSS}.
		(a) The ordering of wires is changed.
		(b) The swap gate is decomposed into a sequence of three CNOT-gates.
		(c) The last CNOT-gate is erased by Eq.~(\ref{eq:measCNOT}).
		(d) Shifting the position of the short $\CZ$-gate (represented by a dashed line) forward in time. The longest $\CZ$-gate is canceled, when the short one passes through the target part of the CNOT-gate. The last CNOT-gate (represented by a dashed line) is decomposed into a $\CZ$-gate and two Hadamard gates.
		(e) Inserting a CNOT-gate by Eq.~(\ref{eq:stateCNOT}) at the positions pointed by the vectors.
		(f) Shifting the inserted CNOT-gate backwards until its target part touches the state $| + \rangle$ and erased by Eq.~(\ref{eq:stateCNOT}). The last $\CZ$-gate is canceled when the CNOT-gate passes through the short $\CZ$-gate. Two Hadamard gates also cancel each other after the $\CZ$-gate is canceled.
		(g) Shifting the position of the CNOT-gate until it collides with the post-selected state $|+ \rangle$ and erased by Eq.~(\ref{eq:measCNOT}).}
	\label{fig:BSS_acausal_transform}
\end{figure*}

The circuit presented in FIG.~\ref{fig:acausal_CZ_BSS} is transformed into our definition of an acausal $\CZ$-gate given in FIG.~\ref{fig:acausal_CZ_def}.
We again use Eqs.~(\ref{eq:stateCNOT}) and (\ref{eq:measCNOT}) and describe the transformation shown in FIG.~\ref{fig:BSS_acausal_transform}.
\newpage
\begin{figure}
	\centering
	\includegraphics[width=6cm]{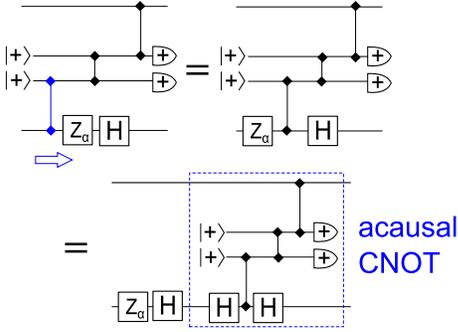}
	\caption{(color online) A proof of the circuit identity presented in FIG.~\ref{fig:acausal_CZ_identity}(b). $\J$-gate $\J_{\alpha}$ is decomposed into a phase gate $Z_{\alpha}$ and Hadamard gate $H$. }
	\label{fig:pf_acCZ_J}
\end{figure}
\begin{figure*}
	\centering
	\includegraphics[width=12cm]{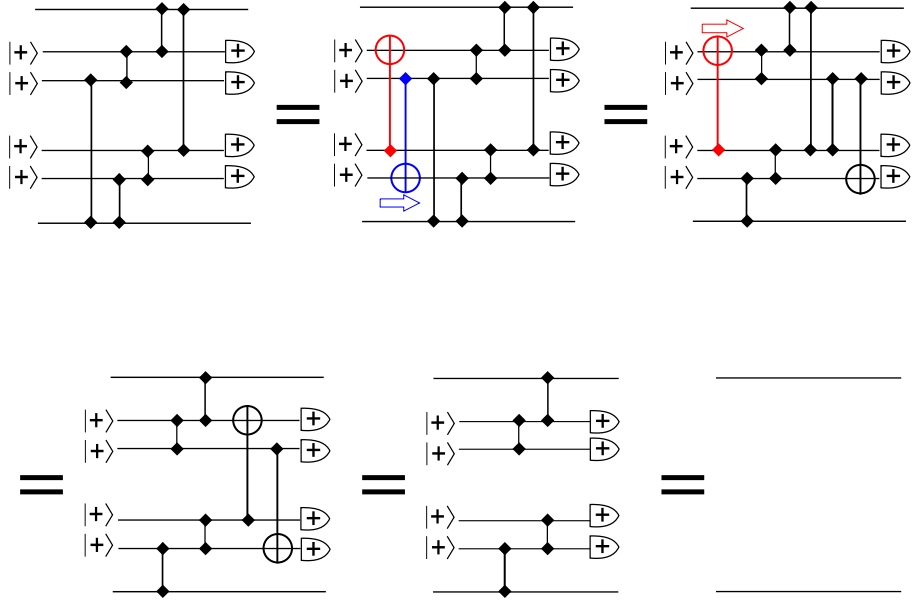}
	\caption{(color online) A proof of the circuit identity presented in FIG.~\ref{fig:acausal_CZ_identity}(d). The first equality is by Eq.~(\ref{eq:stateCNOT}) and the last equality is by Eq.~(\ref{eq:measCNOT}). The second and the third equalities are by Eq.~(\ref{eq:CNOT_CZ}). These circuit translations can be carried out even if the acausal $\CZ$-gates act on a single wire.}
	\label{fig:pf_ac_cancel}
\end{figure*}
\begin{figure*}
	\centering
	\includegraphics[width=12cm]{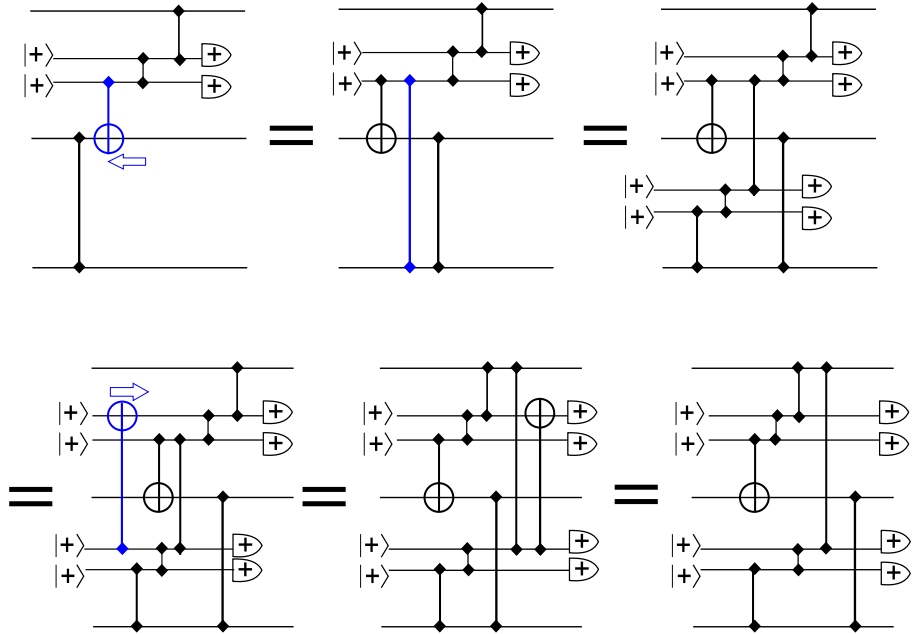}
	\caption{(color online) Preparation for a proof of the circuit identity presented in FIG.~\ref{fig:acausal_CZ_identity}(c). The first and the fourth equalities are by Eq.~(\ref{eq:CNOT_CZ}). The third equality is by Eq.~(\ref{eq:stateCNOT}) and the last equality is by Eq.~(\ref{eq:measCNOT}). An ordinary $\CZ$-gate is changed into an acausal $\CZ$-gate at the second equality.}
	\label{fig:pf_acCNOT_CZ}
\end{figure*}
\begin{figure*}
	\centering
	\includegraphics[width=12cm]{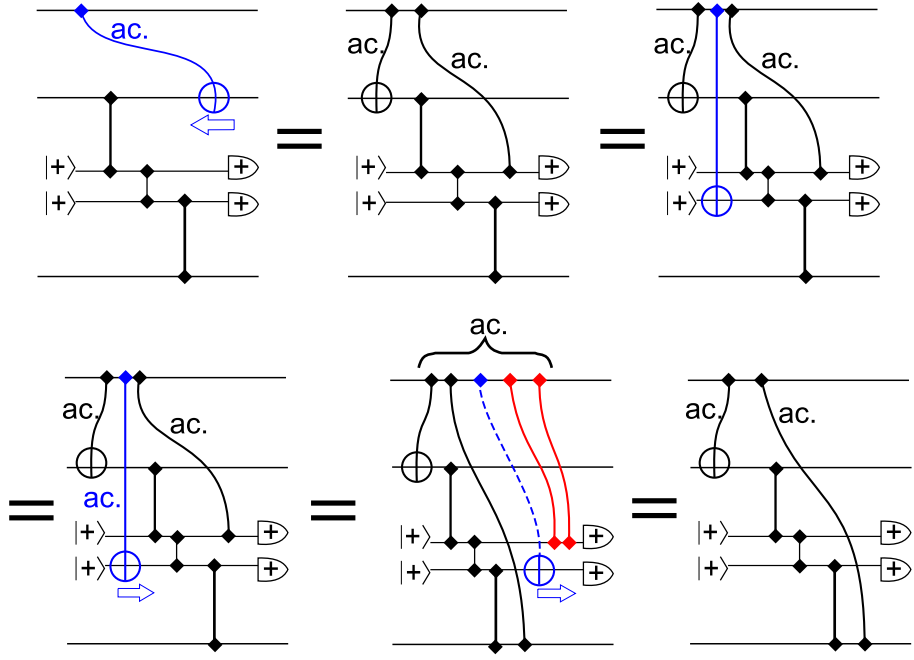}
	\caption{(color online) A proof of the circuit identity presented in FIG.~\ref{fig:acausal_CZ_identity}(c). The first and the fourth equalities are by the circuit identity proved in FIG.~\ref{fig:pf_acCNOT_CZ}. The second equality is by Eq.~(\ref{eq:stateCNOT}) and the dashed acausal CNOT-gate is erased by Eq.~(\ref{eq:measCNOT}) in the last equality. An ordinary CNOT-gate is changed into an acausal CNOT-gate at the third equality. Two acausal $\CZ$-gates appearing on the top right of the second last circuit are canceled by the circuit identity presented by FIG.~\ref{fig:acausal_CZ_identity}(d) in the last equality.}
	\label{fig:pf_acCNOT_acCZ}
\end{figure*}

\section{Controlled stabilizer}\label{sec:append_stabi}
We present a proof of Eq.~(\ref{eq:controlled stabilizer}): the controlled operator $\tCK(g_V(v_i))_{h(v_i);W(v_i)}$ of $\tK(g_V (v_i))$ defined by Eq.~(\ref{eq:matching gflow stabilizer}) also stabilizes the open graph state $| G' \rangle_{\phi}$.

First we prove that the operator $\tK(g_V (v_i))$ ($v_i \in \Vp$) defined by
\begin{eqnarray}
	\tK(g_V(v_i)) &=& \tX_{g_V(v_i) \oplus h(v_i)} \tZ_{Odd_{G'} \left( g_V (v_i) \oplus h(v_i) \right)}
	\label{eq:control stabilizer2}
\end{eqnarray}
is a stabilizer of $| G' \rangle_\phi$.
From the anti-commutation relation $XZ+ZX=0$,
\begin{eqnarray}
\nonumber	(-1)^N \tK(g_V(v_i)) = \prod_{u \in g_V(v_i) \oplus h(v_i)} \tX_u \tZ_{N(u)},
\end{eqnarray}
where
\begin{eqnarray}
N = | \{ (v_1 , v_2) \in E |~ v_1,v_2 \in g(u) \} |
\nonumber
\end{eqnarray}
holds and the right hand side is a stabilizer of $| G' \rangle_\phi$.
Since there are no edges between any pair of vertices in $g_V(v_i) \oplus h(v_i)$ on $G'$, $N=0$ in this case.
It follows that $\tK(g_V(v_i))$ is a stabilizer of $| G' \rangle_\phi$.

Let us exchange the order of $\tCN$ included in $\tCK(g_V(v_i))_{h(v_i);W(v_i)}$ and $\tCP$ constructing the open graph state, by adding extra $\CZ$-gates in a way presented in FIG.~\ref{fig:circuit_identity}(c).
For any $\tCN_{h(v_i); u}$ ($u \in g_V(v_i) \oplus h(v_i)$),
\begin{eqnarray}
\nonumber	\tCN_{h(v_i); u} \tE_{G'} &=& \tE_{G'} \tCP_{h(v_i); N_{G'}(u)} \tCN_{h(v_i); u} \\
\nonumber	&=& \tCP_{h(v_i); N_{G'}(u)} \tE_{G'} \tCN_{h(v_i); u},
\end{eqnarray}
holds.
Summing up for all $u \in g_V(v_i) \oplus h(v_i)$, we have
\begin{eqnarray}
\nonumber	&& \tCN_{h(v_i); g_V(v_i) \oplus h(v_i)} \tE_{G'} \\
\nonumber &=& \prod_{u \in g_V(v_i) \oplus h(v_i)} \tCP_{h(v_i); N_{G'}(u)}  \tE_{G'} \tCN_{h(v_i); g_V(v_i) \oplus h(v_i)} \\
\nonumber	&=& \tCP_{h(v_i);Odd_{G'} \left( g_V (v_i) \oplus h(v_i) \right) }  \tE_{G'} \tCN_{h(v_i); g_V(v_i) \oplus h(v_i)}.
\end{eqnarray}
The newly created $\CZ$-gates are identical to those appearing in $\tCK(g_V(v_i))_{h(v_i);W(v_i)}$.
Thus if we apply $\tCK(g_V(v_i))_{h(v_i);W(v_i)}$ on $| G' \rangle_\phi = \tE_{G'} |+ \rangle_{I^C} |\phi \rangle_I$, it yields
\begin{eqnarray}
\nonumber	\tCK(g_V(v_i))_{h(v_i);W(v_i)} \tE_{G'} |+ \rangle_{I^C}  |\phi \rangle_I \\
	= \tE_{G'} \tCN_{h(v_i); g_V(v_i) \oplus h(v_i)} |+ \rangle_{I^C}  |\phi \rangle_I,
	\label{eq:append3}
\end{eqnarray}
where $\CZ$-gates are canceled.
Since $g_V(v_i) \oplus h(v_i) \cap I = \emptyset$, the target side of CNOT-gates in Eq.~(\ref{eq:append3}) all act on $|+ \rangle$ states.
From Eq.~(\ref{eq:stateCNOT}), these CNOT-gates are eliminated to derive
\begin{eqnarray}
\nonumber	\tCK(g_V(v_i))_{h(v_i);W(v_i)} |G' \rangle_\phi = |G' \rangle_\phi.
\end{eqnarray}


\begin{thebibliography}{99}
	\bibitem{MBQCPRL} R. Raussendorf and H. J. Briegel, Phys. Rev. Lett. {\bf 86}, 5188 (2001).
	\bibitem{graph-states-PRA} M. Hein, J. Eisert, and H. J. Briegel, Phys. Rev. A {\bf 69}, 062311 (2004).
	\bibitem{graph-states-Review} M. Hein, W. D\"ur, J. Eisert, R. Raussendorf, M. Van den Nest, and H. J. Briegel, in {\it Proceedings of the International School of Physics ``Enrico Fermi'' on ``Quantum Computers, Algolithms and Chaos,''} edited by G. Casati {\it et al}. (IOS Press, Amsterdam, 2006), Vol. 162, e-print \eprint{arXiv:quant-ph/0602096}.
	\bibitem{QFT} R. Cleve and J. Watrous, in {\it Foundation of Computer Science, 2000. Proceedings. 41st Annual Symposium on}, pp. 526-536, (2000).
	\bibitem{complexity-of-MBQC} D. E. Browne, E. Kashefi, and S. Perdrix, in {\it Proceeding of the fifth Conference on the Theory of Quantum Computation, Communication and Cryptography} ({\it TQC 2010}), {\it Leeds} 2010.
	\bibitem{Fanout} P.  H{\o}yer and R. {\v S}palek, Theory of Computing {\bf 1}, 81, (2005).
	\bibitem{MBQCPRA} R. Raussendorf, D. E. Browne, and H. J. Briegel, Phys. Rev. A {\bf 68}, 022312 (2003).
	\bibitem{graph-change-by-measurements} M. Mhalla and S. Perdrix, e-print \eprint{arXiv:quant-ph/0412071} (2004).
	\bibitem{flow} V. Danos and E. Kashefi, Phys. Rev. A {\bf 74}, 052310 (2006).
	\bibitem{gflow} D. E. Browne, E. Kashefi, M. Mhalla, and S. Perdrix, New. J. Phys. {\bf 9}, 250 (2007).
	\bibitem{category} R. Duncan and S. Perdrix, in {\it Proceedings of the 37th International Colloquium, ICALP 2010, Bordeaux, Part II,} edited by S. Abramsky {\it et al}., Lecture Notes in Computer Science Vol. 6199 (Springer, Berlin, 2010) pp. 285-296.
	\bibitem{compact} R. Dias da Silva and Ernesto F. Galv\~ao, Phys. Rev. A {\bf 88}, 012319 (2013).
	\bibitem{PCTCcompt} S. Lloyd, L. Maccone, R. Garcia-Patron, V. Giovannetti, and Y. Shikano, Phys. Rev. D {\bf 84}, 025007 (2011).
	\bibitem{DCTCcompt} S. Aaronson and J. Watrous, Proc. R. Soc. A {\bf 465}, 631 (2009).
	\bibitem{vicious-circuit} N. de Beaudrap, Phys. Rev. A {\bf 77}, 022328 (2008).
	\bibitem{Diestel} R. Diestel, {\it Graph Theory} ({\it Graduate Texts in Mathematics}) (Springer, 2005).
	\bibitem{flow-to-circuit} A. Broadbent and E. Kashefi, Theor. Comput. Sci. {\bf 410}, 26 (2009).
	\bibitem{max-delayed-gflow} M. Mhalla and S. Perdrix, in {\it Automata, Languages and Programming} {\bf 5125}, edited by L. Aceto {\it et al}. (Springer Berlin Heidelberg, 2008) p. 857.
	\bibitem{global-circuit-optimization} R. Dias da Silva, E. Pius, and E. Kashfi, e-print \eprint{arXiv:1301.0351} (2013).
	\bibitem{simulating-CTC-by-MBQC} R. Dias da Silva, Ernesto F. Galv\~ao, and E. Kashefi, Phys. Rev. A {\bf 83}, 012316 (2011).
	\bibitem{CTCBS} C. H. Bennett and B. Schumacher (unpublished). See also \url{http://web.archive.org/web/20070206131550/http://www.research.ibm.com/people/b/bennetc/QUPONBshort.pdf}.
	\bibitem{CTCS} G. Svetlichny. e-print \eprint{arXiv:0902.4898} (2009).
	\bibitem{gate-teleportation} D. Gottesman and I. L. Chuang, Nature {\bf 402}, 390 (1999).
\end{thebibliography}

\end{document}